\documentclass[preprint,tightenlines,nofootinbib,aps,showpacs]{revtex4}

\usepackage{bm}% bold math
\usepackage[utf8]{inputenc}
\usepackage{graphicx}
\usepackage[colorlinks]{hyperref}
\usepackage[utf8]{inputenc}
\usepackage{amsmath}
\usepackage{amssymb}
\usepackage{slashed}

\def\nn{\nonumber}

\begin{document}

%============================================================
% Title Page
%============================================================

\preprint{OU-HEP-220208}

% \title{\boldmath
% Searching for extra Higgs bosons 
% via $pp\to H,A\to \tau\mu, \tau\tau $ \\
% at the Large Hadron Collider}

\title{\boldmath
Prospects for extra Higgs boson search 
via $pp\to H,A\to \tau\mu, \tau\tau $ \\
 at the High Luminosity Large Hadron Collider
}

\author{Wei-Shu Hou}
\email{wshou@phys.ntu.edu.tw}
 \affiliation{Department of Physics, National Taiwan University,\ Taipei 10617, Taiwan}
\author{Rishabh Jain}%
 \email{rishu25.RJ@gmail.com}
\affiliation{%
 Department of Physics, National Taiwan University,\ Taipei 10617, Taiwan}%
\author{Chung Kao}%
 \email{chung.kao@ou.edu}
\affiliation{%
 Homer.L. Dodge Department of Physics and Astronomy, 
 University of Oklahoma,
 Norman, Oklahoma 73019, USA}%

\bigskip

\date{\today}

\bigskip

%-----------------------------------
%   Abstract
%-----------------------------------

\begin{abstract}
We extend heavy Higgs searches at the Large Hadron Collider for 
$H \to \tau \mu$ by CMS, and $H \to \tau\tau$ by ATLAS and CMS, 
to study discovery prospects of extra Higgs states in $pp \to H,A \to 
\tau \mu, \tau\tau$ with $e\mu + \slashed{E}_T$ and $j_{\tau} \mu 
+ \slashed{E}_T$ final states, where $j_{\tau} = \pi \, , \rho\, , a_1$ 
and $\slashed{E}_T$ is missing transverse energy. 
In a general two Higgs doublet model without $Z_2$ symmetry, 
extra Yukawa couplings $\rho_{\tau\tau}$ and $\rho_{\tau\mu}$ can 
drive $H,A \to \tau\tau$ and $\tau\mu$ channels at hadron colliders, 
following gluon-gluon fusion production with extra $\rho_{tt}$ couplings. 
The light Higgs boson $h(125)$ is found to resemble closely the Standard 
Model Higgs boson; in the alignment limit of $\cos\gamma \to 0$ for 
$h$--$H$ mixing, flavor changing neutral Higgs couplings such as 
$h \to \tau\mu$ are naturally suppressed, but the couplings of the 
heavier $H$ is optimized by $\sin \gamma \to 1$. We define various 
signal regions for $H,A \to \tau\mu$ and $\tau \tau$ and evaluate physics 
backgrounds from dominant processes with realistic acceptance cuts and 
tagging efficiencies. Two different studies are presented. We first perform 
a parton level study without any hadronization and with minimal detector 
smearing. We then include hadronization using PYTHIA 8.2 and fast 
detector simulation using DELPHES to give event level simulation. 
Results for $\sqrt{s}= 13$ TeV appear promising, which we extend 
further to $\sqrt{s} = 14$ TeV for the High Luminosity LHC.
\end{abstract} 

\maketitle

%-------------------------------------------------------
% Section I Introduction
%-------------------------------------------------------
\section{Introduction}

A 125 GeV scalar, $h$, was discovered in 2012 
by ATLAS and CMS~\cite{ATLAS_SMHiggs,CMS_SMHiggs}, 
which resembles the Higgs boson of the Standard Model (SM), 
marking the success of SM up to the electroweak scale. 
Despite the remarkable resemblance of $h$ with the SM Higgs, 
we are still unclear about the nature of 
electroweak symmetry breaking (EWSB), and 
whether the SM Higgs sector is complete remains a mystery. 
Many extensions of SM expand the Higgs sector by 
adding additional doublets~\cite{Branco:2011iw} 
like two Higgs doublet models (2HDM), 
minimal SUSY (MSSM), 
three Higgs doublet model (3HDM)~\cite{Keus:2013hya}.

In this article, we study one of the simpler 
extensions of SM, a.k.a. 2HDM, where we extend 
the SM Higgs sector by an additional Higgs doublet, 
with both doublets coupling to fermions. 
As a result, we have two Yukawa matrices that 
cannot be simultaneously diagonalized, and thus 
we have off-diagonal flavor violating terms. 
The off-diagonal Yukawa terms can generate tree level 
flavor-changing neutral Higgs (FCNH) interactions; 
this version is referred to as 
2HDM-III~\cite{Hou:1991un} or general 2HDM (g2HDM). 
The FCNH interactions are usually avoided by introducing 
some {\it ad hoc} $Z_2$ symmetries to enforce
the Glashow-Weinberg natural flavor conservation (NFC) condition~\cite{Glashow:1976nt}, giving rise to 
2HDM Type I, II, X, and Y versions~\cite{Branco:2011iw}. 
However, we do not enforce $Z_2$ symmetries 
but let {\it nature} provide us with its true flavor.

Recently, the Fermilab 
Muon g-2 experiment~\cite{Muong-2:2021ojo} confirmed 
the previous result of the BNL 
Muon g-2 experiment. Their combined result is,
\begin{equation}
 a_{\mu}(\rm Exp) = 116592061(41)
               \times 10^{-11} (0.35 \, \rm ppm),
\end{equation}
which deviates from the community consensus 
SM expectation~\cite{Aoyama:2020ynm}, 
$a_{\mu}(\rm SM) = 116591810 (43)
              \times 10^{-11} (0.37 \, \rm ppm)$, 
by 4.2$\sigma$. The muon $g-2$ anomaly 
can be explained in g2HDM by flavor violating 
$\rho_{\tau\mu}$ couplings, as discussed 
in Refs.~\cite{Wang:2021fkn,Hou:2021sfl,Athron:2021jyv}. 
%the authors presented a possibility of explaining the $g-2$ anomaly by setting $\rho_{\tau\mu} \sim 20-30 \lambda_{\tau}\,(\lambda_{\tau} \sim 0.01)$, i.e the muon $g-2$ excess could be a hint of flavor violation in the leptonic sector. 
%
The lepton flavor violating (LFV) $\rho_{\tau\mu}$ 
and $\rho_{\mu\tau}$ couplings can drive $h \to \tau \mu$, 
which can be of concern as the limit~\cite{CMS:htamu}
\begin{equation}
    \mathcal{B}(h \to \tau \mu) < 0.15\, \%,
\end{equation}
is rather stringent. However, one can overcome 
the strong bounds with the help of 
{\it Alignment}~\cite{Hou:2017hiw}. 
Under alignment the properties of $h$ 
closely resembles that of SM Higgs, 
which requires the mixing angle between 
the two CP-even scalars $h$,\,$H$ to
approach zero, $\cos\gamma \to 0$, with the 
$h\tau\mu$ coupling $\propto \rho_{\tau\mu}\cos\gamma$. 

In g2HDM, the exotic scalar $H$ benefit from alignment 
with $\sin\gamma \to 1$, and there is no suppression for 
$H \to \tau \mu$ or $A \to \tau \mu$ LFV processes.
This property was exploited in Ref.~\cite{Hou:2019grj,Arganda:2019gnv}, 
where a detailed collider search was performed. 
Subsequently, CMS published~\cite{CMS:2019pex} 
a detailed search for $H \to \tau \mu$ for 
$m_H \in [200,\,900]$\,GeV with 35.9\;fb$^{-1}$ data. 
No excess was found, placing strong limits on 
the $gg \to H \to \tau \mu$ cross section, 
but CMS has yet to update with full Run 2 data. 
The $\rho_{\tau\mu}$, $\rho_{\mu\tau}$ couplings 
along with $\rho_{tt}$ also contribute 
to $\tau \to \mu \gamma$ via two-loop 
Bjorken-Weinberg (or Barr-Zee) mechanism, 
which dominates over the one-loop mechanism, 
provided that $\rho_{tt} \sim 
\mathcal{O}(\lambda_t)$~\cite{Chang:1993kw,Hou:2020tgl}. 
The one-loop effect would be suppressed 
if one takes~\cite{Hou:2020itz}
$\rho_{\tau\tau} = \rho_{\tau\mu} = \rho_{\mu\tau}
 = \lambda_{\tau} \sim \mathcal{O}(0.01)$. 
%especially when the mass difference 
%between $H$ and $A$ is large, as considered in this article. 
We further extend our work from Ref.~\cite{Hou:2019grj} by respecting the current limits on $gg \to H\to \tau \mu$ cross-sections and $\mathcal{B}(\tau \to \mu\gamma)$.

The extra $\tau$ Yukawa coupling $\rho_{\tau\tau}$ 
with alignment is the main driver for 
the $gg \to H,A \to \tau\tau$ process. 
In addition, $\rho_{\tau\tau}$ can carry 
a {\it complex} phase, which can contribute to 
$\tau$ electric dipole moment~\cite{Hou:2021zqq}, 
or reveal itself in the CP structure of 
the $h\tau\tau$ coupling. The complex phase of 
the $h\tau\tau $ coupling is extensively searched 
by ATLAS~\cite{ATLAS:2020evk} and CMS~\cite{CMS:2021sdq}. 
In addition, 
CMS~\cite{CMS:HTATA} and ATLAS~\cite{ATLAS:HTATA} 
have also searched for the heavy exotic scalars 
decaying to $\tau\tau$ in the 200--2500~GeV mass range. 
This motivates us to study the collider prospects for 
$H,A \to \tau \tau$ and provide predictions for HL-LHC. 

This article is organized as follows. 
We first briefly review g2HDM in Sec.~II, 
and derive in Sec.~III the constraints from 
the experiments on important couplings 
relevant to our collider study. Sec.~IV discusses the prospects of $p p\to H,A \to \tau\mu +X$, 
while Sec.~V is reserved for $p p \to H,A \to \tau \tau +X$. 
We present the discovery potential of 
both $\tau\mu$ and $\tau\tau$ in Sec.~VI 
and conclude in Sec.~VII.

%--------------------------------------------------------------
%  Frame work  Section II
%--------------------------------------------------------------
\section{The general two Higgs doublet model}

In g2HDM, one can write the Higgs potential in 
the Higgs basis, namely~\cite{Hou:2017hiw,Davidson:2005cw}
\begin{eqnarray}
 V(\Phi,\Phi^{'}) & = & \mu_{11}^2 |\Phi|^2 + \mu_{22}^2 |\Phi'|^{2} - (\mu_{12}^2 \Phi^{\dagger} \Phi' + h.c)  
  + \frac{1}{2}\eta_1 |\Phi|^4 + \frac{1}{2}\eta_2 |\Phi'|^4 \nn \\
& + &   \eta_3 |\Phi|^2|\Phi'|^2 + \eta_4 |\Phi^{\dagger}\Phi'|^2   
  +  \left[\Bigl (\frac{1}{2}\eta_5 \Phi^{\dagger}\Phi'
                   + \eta_6 |\Phi|^2 + \eta_7 |\Phi'|^2\Bigr)\Phi^{\dagger}\Phi' + h.c.\right] ,
\label{eq:Pot}
\end{eqnarray}
where EWSB arises from $\Phi$ while 
$\langle\Phi'\rangle = 0$ (hence $\mu_{22}^2 > 0$).
In Eq.~(\ref{eq:Pot}), $\eta_i$s are 
the quartic couplings and taken as real,
as we assume the Higgs potential is CP-invariant. 
After EWSB, one can find~\cite{Hou:2017hiw} from Eq.~(\ref{eq:Pot})
the mass eigenstates $h$, $H$, $A$ and $H^+$, 
as well as $h$-$H$ mixing, 
where we define the mixing angle as $\gamma$.
In the alignment limit, $\cos\gamma \to 0$.

The Yukawa couplings of the Higgs bosons to quarks
are given as~\cite{Davidson:2005cw,Chen:2013qta},
\begin{eqnarray}
 -\frac{1}{\sqrt{2}}\sum_{f = u,d}
&  \bar{f}_i\left[
            (\lambda_{i}^f\delta_{ij}s_{\gamma}
                      + \rho^{f}_{ij}c_{\gamma})h
          - (\lambda_{i}^f\delta_{ij}c_{\gamma}
                      - \rho^{f}_{ij}s_{\gamma})H
          - i\,\mathrm{sgn}(\mathcal{Q}_f)\rho^{f}_{ij}A
            \right] R f_{j} \nn \\
& -\bar{u}_{i}\bigl[(V\rho^d)_{ij} R
                  - (\rho^{u\dagger}V)_{ij} L
              \bigr] d_{j}H^{+}
  + {\rm h.c.},
\label{eq:Yuk}
\end{eqnarray}
where $\lambda_i^f = \sqrt2 m_f/v$ is the SM Yukawa coupling, 
$\rho^f$ is the extra Yukawa matrix and 
$c_{\gamma}(s_{\gamma}) \equiv \cos\gamma (\sin\gamma)$. 
An analogous equation holds for charged leptons, 
but with $V$ set to unity because of 
the rather degenerate neutrinos.
As discussed in the Introduction, $\rho^f$ can carry 
nonzero off-diagonal flavor violating terms. 
From Eq.~(\ref{eq:Yuk}), the extra Yukawa matrix $\rho^f$ 
is combined with $c_{\gamma}$ for $h$, 
hence the $hf\bar{f}$ coupling vanishes 
in the alignment limit of $c_\gamma \to 0$.  
As a result, all LFV processes such as $h \to \tau\mu$ 
as well as $t \to c h$ are highly suppressed in g2HDM. 
On the upside, $c_\gamma \to 0$ implies\footnote{
  $A$ and $H^+$ couplings are unaffected by alignment,
  as is evident from Eq.~(\ref{eq:Yuk}).} 
$s_\gamma \to 1$, and nonzero flavor violating couplings 
like $\rho_{tc}$,\, $\rho_{\tau\mu}$ can drive 
our signal $gg \to H,A \to \tau \mu$ processes, 
or $gg \to H,A \to tc$~\cite{Altunkaynak:2015twa}, 
even process like $cg \to tH, tA \to 
tt\bar{c}$~\cite{Kohda:2017fkn,Gori:2017tvg}, 
$cg \to tH,tA \to t\tau\mu$ 
and $cg \to bH^+\to bt\bar{b}$~\cite{Ghosh:2019exx}, 
$bHW^+ (\to b\tau\mu W^+, btcW^+$~\cite{Hou:2021sfl}). 
We hence see that, even in the alignment limit, 
g2HDM can provide a rich phenomenology at the LHC. 

The $\rho_{\tau\tau}$ coupling in the alignment limit 
is one of the drivers for our second signal process, 
$pp \to H,A \to \tau \tau$. 
In addition, with complex $\rho_{\tau\tau}$, 
%with phase $\phi_{\tau\tau}$, 
the $h\tau\tau$ coupling can become complex 
hence CP violating, with the phase~\cite{Hou:2021zqq},
\begin{equation}
    {\tan\phi_{h\tau\tau} =
    \frac{c_{\gamma}\,{\rm Im}\,\rho_{\tau\tau}}
         {c_\gamma\,{\rm Re}\,\rho_{\tau\tau} +
                     s_{\gamma}\,\lambda_{\tau}}.}
\end{equation}
In this article, we do not explore the complexity of $\rho_{\tau\tau}$ and $\rho_{\tau\mu}$ 
but keep them real for simplicity. 
Furthermore, unlike Ref.~\cite{Hou:2021sfl}, 
we follow the ``normal'' or conservative guesstimate~\cite{Hou:2020itz} of 
the associated extra Yukawa couplings,
\begin{equation}
\rho_{\tau\tau} = \rho_{\tau\mu} = \mathcal{O}(\lambda_{\tau}),
\quad\ \rho_{tt} = \mathcal{O}(\lambda_t).
\label{eq:rho_ij}
\end{equation}

\begin{figure*}[t]
\centering
\includegraphics[scale=0.18]{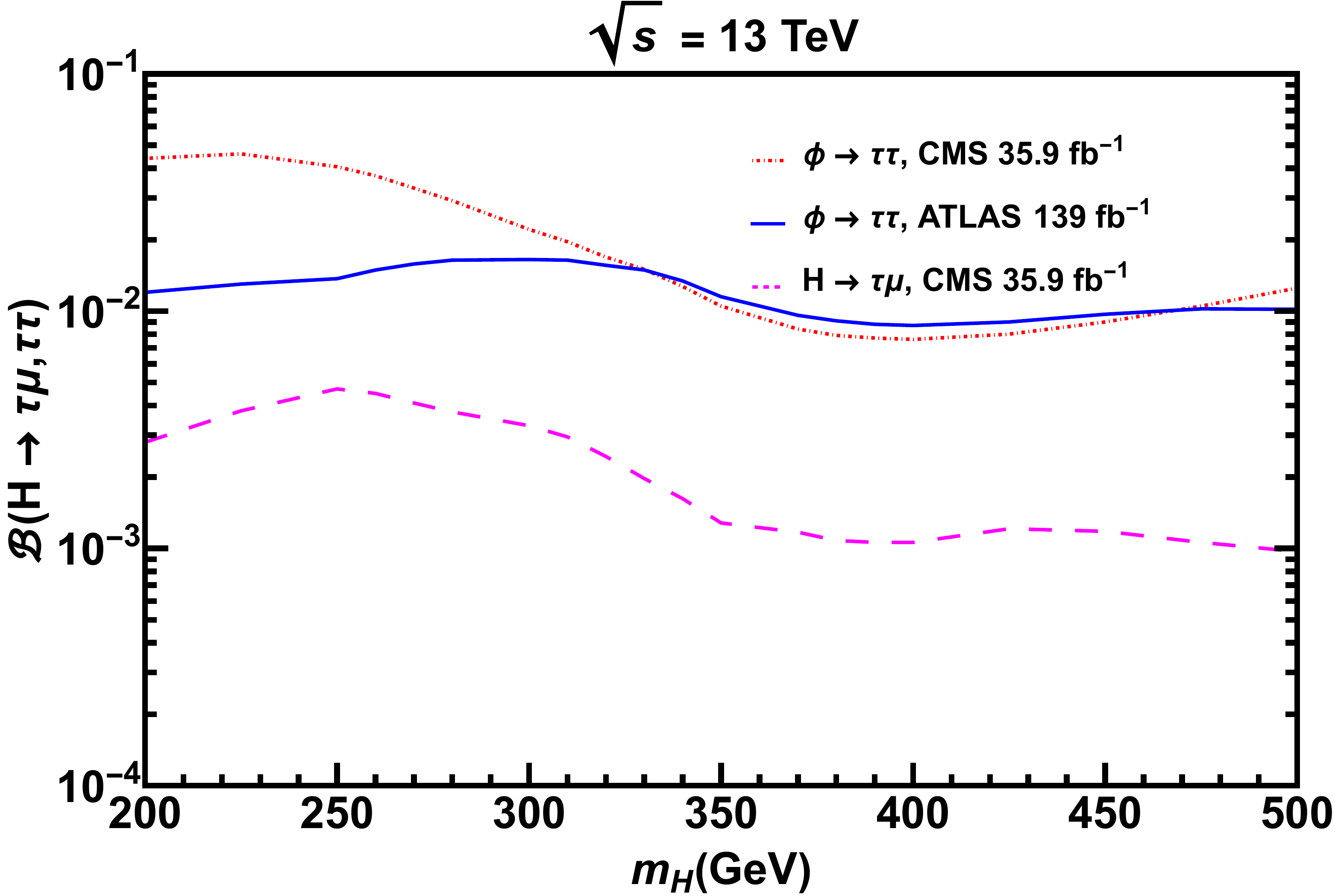}
\caption{
 Limits from CMS and ATLAS on $\mathcal{B}(H\to \tau\mu, \tau\tau)$
 under the assumption of SM-like (exotic) $H$ production.
}
\label{lhc_limit}
\end{figure*}

%\item Main goal of the paper: 
%\begin{itemize}
%\item To present the excellent sensitivity of CMS over ATLAS and Belle experiments.
%\item Collider predictions for current and future LHC
%\item CPV $\to$ complexity ($\rho_{tt}$) predictions and interesting variables to explore. % guessestimate

% \end{itemize}
%\end{itemize}
%The discovery of 125 GeV scalar in 2012 by CMS and ATLAS resembles the predicted SM Higgs, responsible for spontaneous symmetry breaking . However we are still unclear about the electroweak symmetric breaking mechanism as well the flavor structure of nature.    

%-------------------------------------------------------
% Section III Constraints
%-------------------------------------------------------

\section{Constraints on relevant parameters}

The important parameters governing 
$pp \to H,A \to \tau \mu + X $ and 
$pp \to H,A \to \tau\tau + X $ are $\rho_{\tau\mu}$, 
$\rho_{\tau\tau}$, $\rho_{tt}$, $c_{\gamma}$. 
Finite $\rho_{tc}$ can drive 
$H,A \to t\bar{c} + \bar{t}c$~\cite{Altunkaynak:2015twa} 
and dilute $H,A \to \tau\mu, \tau\tau$. 
Note that $\rho_{ct}$ suffers strict 
constraints from $B_s$--$\bar{B}_s$ mixing as it 
enters the process via top-loop~\cite{Crivellin:2013wna}, 
while $\rho_{tc}c_\gamma$ is bound by direct searches 
by CMS and ATLAS for $t \to c h$ decay. 
Recently, CMS~\cite{CMS:2021hug} puts the most stringent bound on 
$\mathcal{B} (t \to ch) < 7.3 \times 10^{-4}$ at 95\% C.L. with $h\to \gamma \gamma$
(this has been recently surpassed by ATLAS~\cite{ATLAS:2023ujo}, 
though at comparable sensitivity). 
Note that the bound depends on $c_{\gamma}$, and in 
the alignment limit ($c_{\gamma} \to 0$) the bound vanishes. 
An interesting point about the 
$t \to c h \to c \gamma \gamma$ channel 
is that both $\rho_{tc}$ through $tch$ and $\rho_{tt}$ 
via $h\gamma\gamma$ enters the decay chain. 
However, in this article we do not explore 
the implications of the CMS $t\to ch$ study on 
the interplay between $\rho_{tt}$ and $\rho_{tc}$ couplings, 
since both effects vanish with alignment. 
Following a simple scaling of~\cite{Aaboud:2018oqm,Jain:2019ebq,Gutierrez:2020eby},
\begin{equation}
    \lambda_{tch} \equiv \rho_{tc} \rm c_{\gamma}
 = 1.92\times\sqrt{\mathcal{B}(t\to ch)}.
 \label{eq:lam_tch}
\end{equation}
we get $\rho_{tc} < $ 0.52 and 5.2 for $c_\gamma$ =0.1, and 0.01, respectively. 

The $\rho_{tt}$ coupling is responsible for 
the production of the exotic scalars by gluon-gluon fusion 
via top-loop, and it is constrained by $B$ physics, 
especially $B_{q}$--$\bar{B}_{q}$ mixing and $b\to s\gamma$, 
as well as direct searches for 
$gg \to \bar{t}H^{+} b \to \bar{t} t \bar{b} b
 + {\rm h.c.}$~\cite{CMS:2020imj,ATLAS:2021upq}. 
We find that~\cite{Hou:2019grj} $B$ physics 
puts stronger bounds than direct searches. 
% on $\rho_{tt}$. 
So we fix (Eq. (7) of Ref.~\cite{Hou:2019grj}) 
\begin{equation}
%  \rho_{tt} = 0.2 \times m_{H^+}\,({\rm GeV})/150.
  \rho_{tt} = 0.2 \times \left( \frac{m_{H^+}}{150 \; {\rm GeV}}
   \right) \, . 
 \label{eq:rho_tt}
\end{equation}
%
%from Ref.~\cite{Hou:2019grj}, where Higgs mass is in GeV units.

In Fig.~\ref{lhc_limit} we present the limits on 
$\mathcal{B} (H \to \tau \mu)$ and 
$\mathcal{B}(H \to \tau\tau)$ assuming 
SM-Like production for simplicity. We do not enforce 
this assumption in the rest of the article. 
An interesting result emerges: 
we find that CMS $H\to \tau\mu$ is 
much more stringent than the ATLAS $H\to \tau\tau$.
% NEW
For the case of $pp \to H \to \tau\mu +X$, the exotic Higgs mass is 
reconstructed with the invariant mass, $M_{\tau\mu}$, using collinear 
approximation in tau decays that allows CMS to put a more
stringent limit than $pp \to \phi \to \tau \tau +X$, in which they 
applied the less precise cluster transverse mass.
For $pp \to H \to \tau\tau +X$ with 330 GeV $\alt m_H \alt$ 470 GeV,
the CMS limit appears mildly better than ATLAS.
However, it is probably within uncertainty.

\begin{figure*}[t]
\includegraphics[scale=0.15]{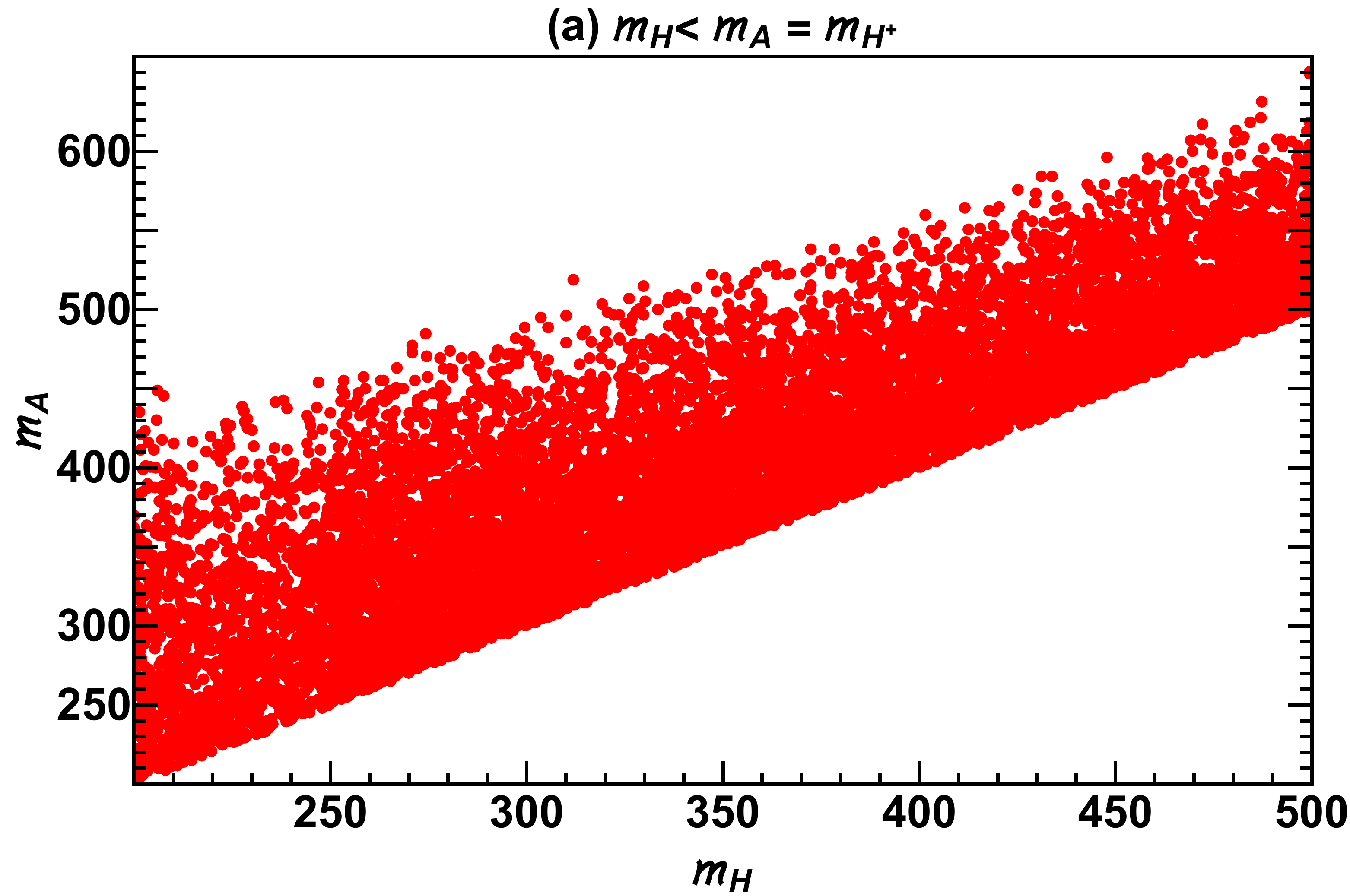}
\includegraphics[scale=0.15]{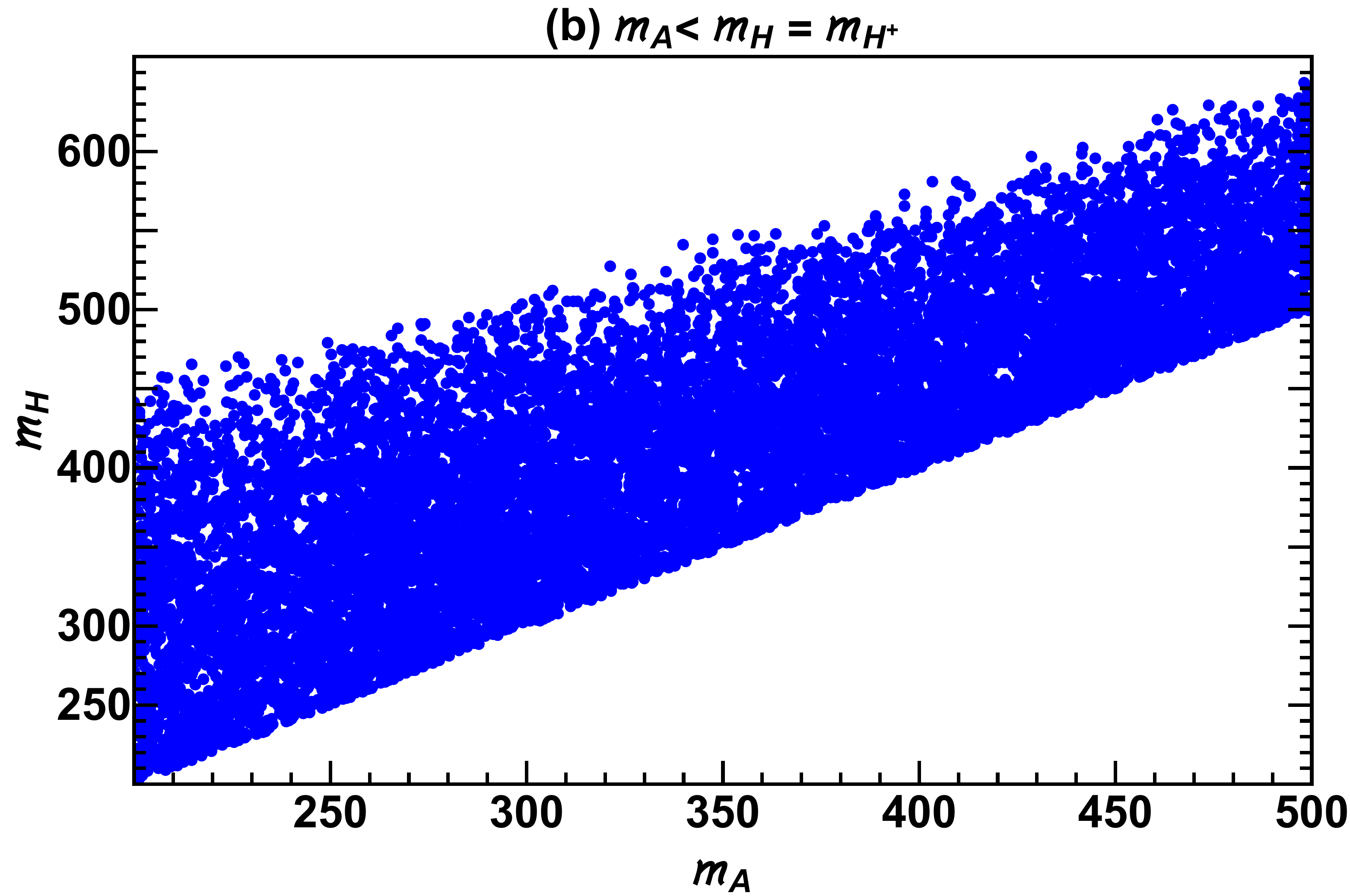}
\caption{
 Scan of allowed points in $m_A$-$m_H$ plane 
 for Cases A and B (see text for more details).
}
\label{Scan_2HDMC}
\end{figure*}

In this article, for simplicity we set all 
off-diagonal $\rho_{ij} = 0$ except $\rho_{\tau\mu}$, 
and all diagonal $\rho_{ii} \sim \lambda_i$ 
except $\rho_{tt}$ and $\rho_{\tau\tau}$. 
The limits on the extra $\tau$ Yukawa couplings $\rho_{\tau\mu}$ and $\rho_{\tau\tau}$ depend on the 
choice of mass and mass differences for $A$, $H$ and $H^+$.
We shall consider four different scenarios, 
\begin{eqnarray}
 {\rm Case\ A1}: && \ m_{H} < m_A = m_{H^+} ,\; \rm  c_{\gamma} = 0.01, \nn \\
 {\rm Case\ A2}: && \ m_{H} < m_A = m_{H^+} ,\; \rm  c_{\gamma} = 0.1, \nn \\
 {\rm Case\ B1}: && \ m_{A} < m_H = m_{H^+} , \; \rm c_{\gamma} = 0.01, \nn \\
 {\rm Case\ B2}: && \ m_{A} < m_H = m_{H^+} ,\; \rm  c_{\gamma} = 0.1.  
 \label{eq:cases}
\end{eqnarray}
%
% NEW: Limits on \cos_\gamma
%
ATLAS~\cite{ATLAS:2019nkf,ATLAS:2021vrm} and
  CMS~\cite{CMS:2018uag} have placed limits on
  $c_\gamma = \cos(\beta-\alpha)$ for 4 types of Yukawa interactions in
  two Higgs doublet models where extra Yukawa couplings are absent:
  Type~I ($|c_\gamma| < 0.3$),
  Types~II, Lepton-Specific, and Flipped ($|c_\gamma| < 0.1$).
With extra Yukawa coupling matrices, however, it would be harder to constrain 
$c_\gamma$. But the fact that $h(125)$ so resembles the SM Higgs boson
 --- alignment, for simplicity, we choose $c_\gamma = 0.1$ and $0.01$ in
  Eq.~(9) as benchmarks for the alignment limit.

We consider the mass difference $|m_H - m_A| =$ 150 GeV. 
A higher mass difference of around 150~GeV 
may run into various constraints.
To show allowed parameter space, 
we perform a random scan 
by setting $m_h = 125.1$\,GeV, $\eta_{2} \in [0, 5]$ 
and $\eta_{7} \in [-5, 5]$, 
and for the lighter $H$ case scan $m_{H} \in [200, 500]$\,GeV 
and $m_{A} \in [200, 700]$\,GeV. 
The scan result is given in Fig.~\ref{Scan_2HDMC}(a), 
and vice versa for the lighter $A$ case in Fig.~\ref{Scan_2HDMC}(b). 
All the points in Fig.~\ref{Scan_2HDMC} 
satisfy {(see e.g. Ref.~\cite{Hou:2021sfl})} 
vacuum stability, perturbativity, 
unitarity and $T$-parameter constraints.

\begin{figure*}[t]
\includegraphics[scale=0.15]{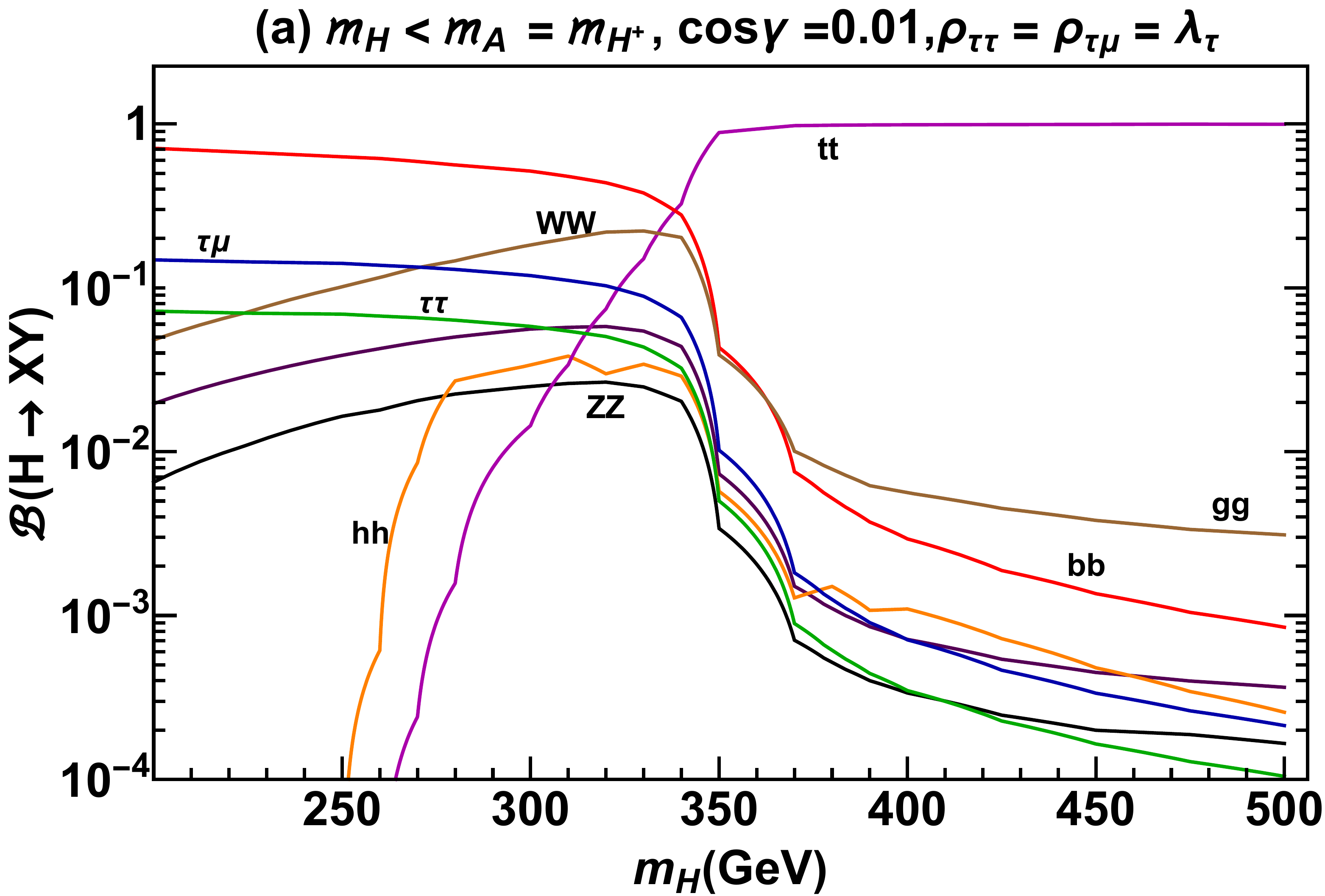}
\includegraphics[scale=0.15]{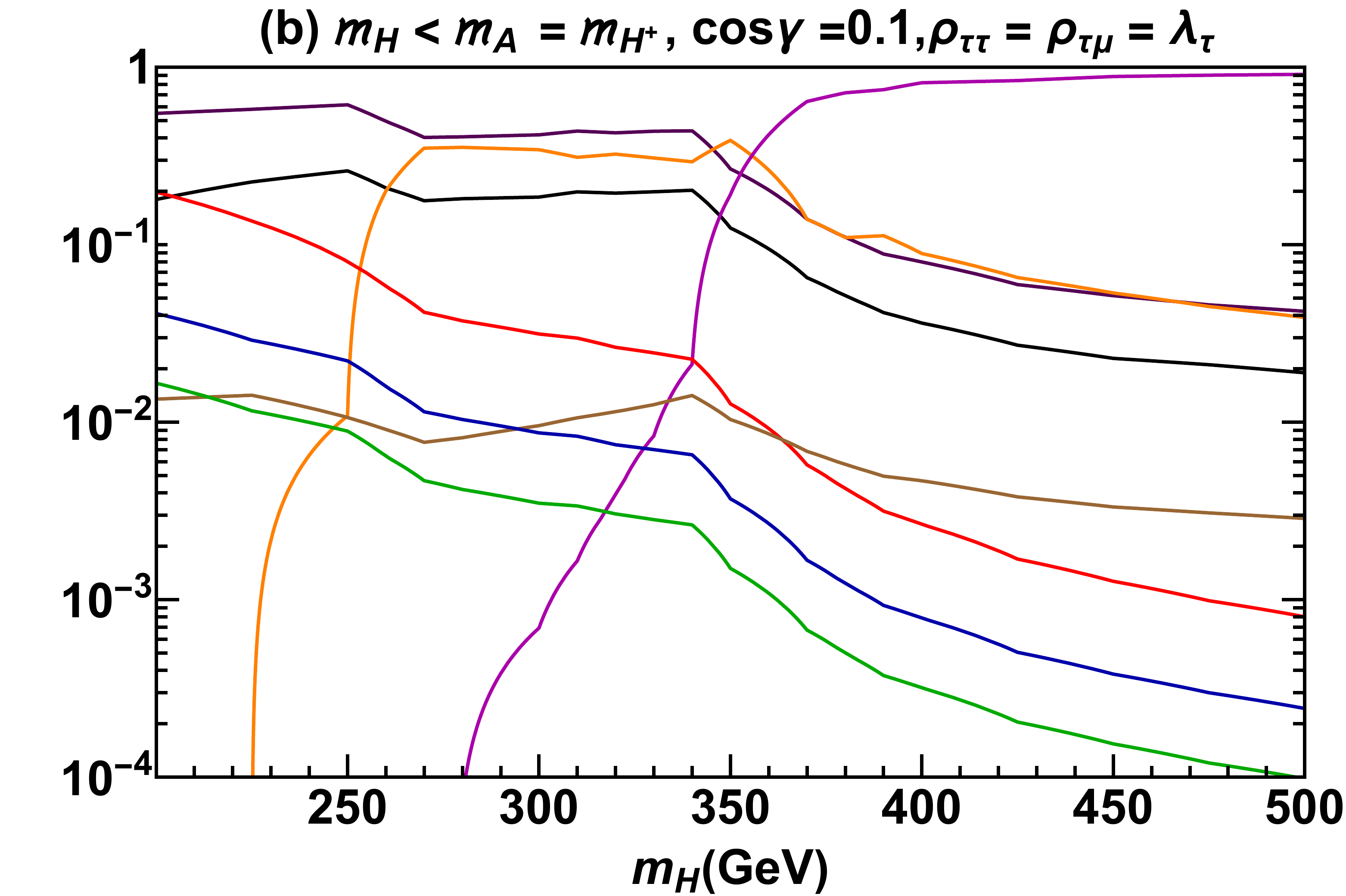} \\
\includegraphics[scale=0.15]{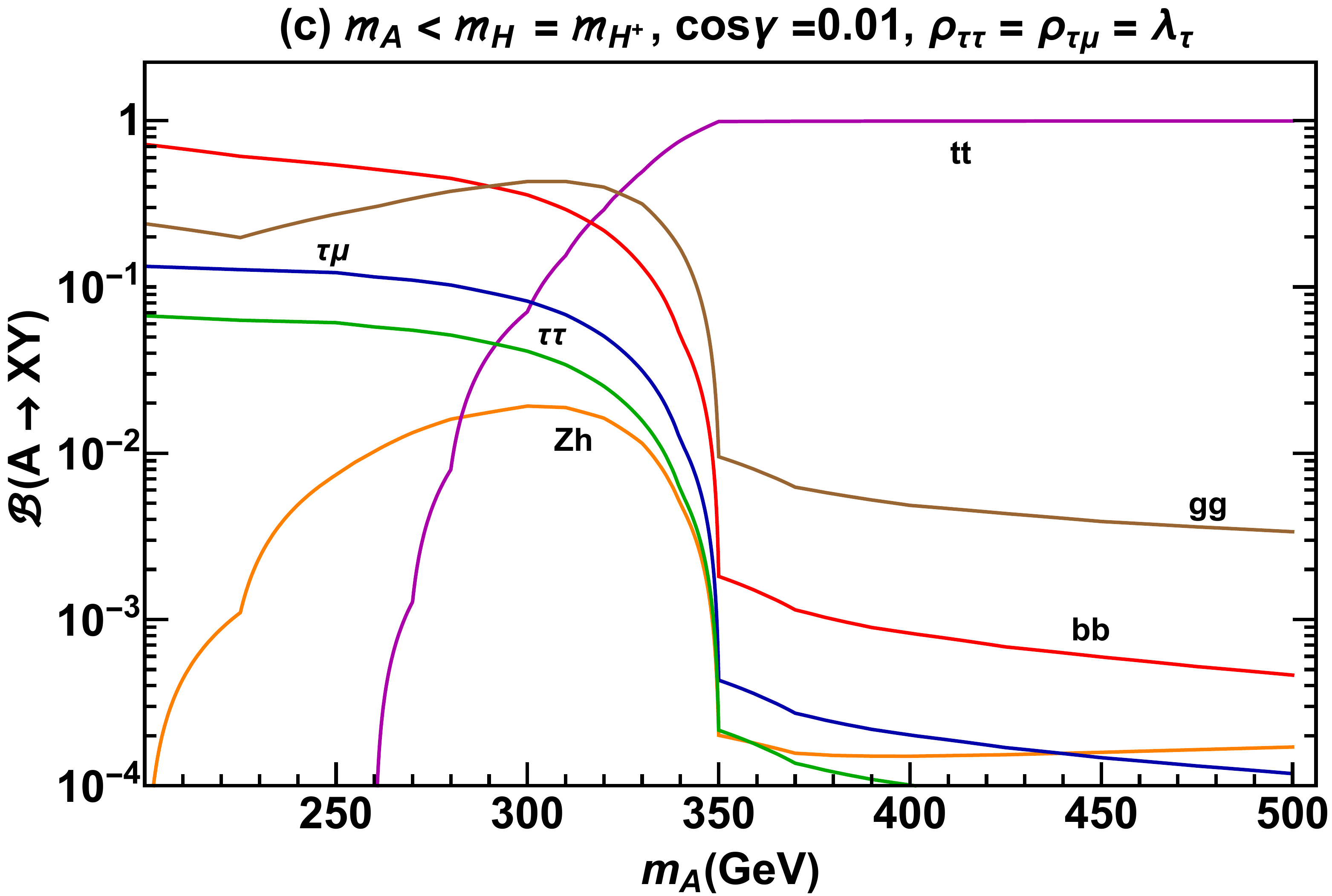}
\includegraphics[scale=0.15]{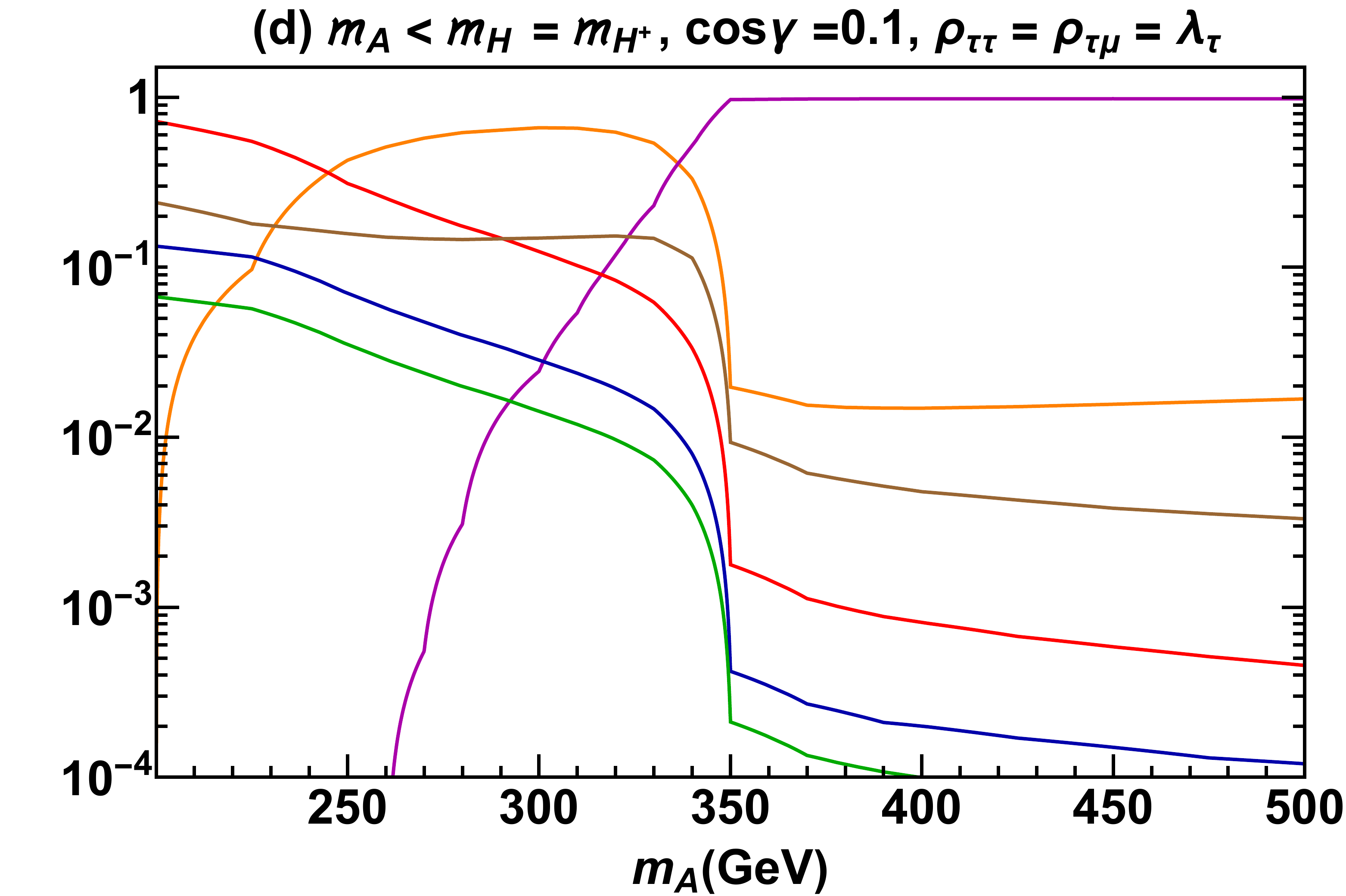} \\
\caption{
  Branching fractions for (a), (b) $H$ decays, 
  and (c), (d) $A$ decays.
}
\label{BF}
\end{figure*}

We see from Fig.~\ref{Scan_2HDMC} that 
the allowed $|m_H - m_A|$ difference decreases 
as we increase the mass of $H$ or $A$. 
We select $m_H = [200, 500]$ GeV for Cases A1 and A2, 
and $m_A = [200, 500]$ GeV for Cases B1 and B2, 
using some random points from scan that satisfy the mass-difference 
to get an estimate of $\lambda_{Hhh}$~\cite{Hou:2019qqi},\footnote{
  Requiring $\lambda_{Hhh}/v < 1.0$.
} 
i.e the trilinear-Higgs coupling for $H\to hh$.\footnote{
  This is only relevant for Cases A1 and A2.
}  
In Fig.~\ref{BF} we present the branching fractions 
for different decay modes of $H$ and $A$ for 
all four cases of Eq.~(\ref{eq:cases}). 
Note that $\rm c_{\gamma}$ does not affect any 
fermionic decay width of $A$, although 
$\Gamma (A \to Z h) \propto  |\rm c_{\gamma}|^2$.  

% these plots needs to have regions, that would be more robust. that would  0.01 < cos\gamma < 0.1, and allowed values of GHhh based on your scan with max value of 50 GeV (80 GeV) , similar arguments with the constraints

\begin{figure*}[t]
\includegraphics[scale=0.15]{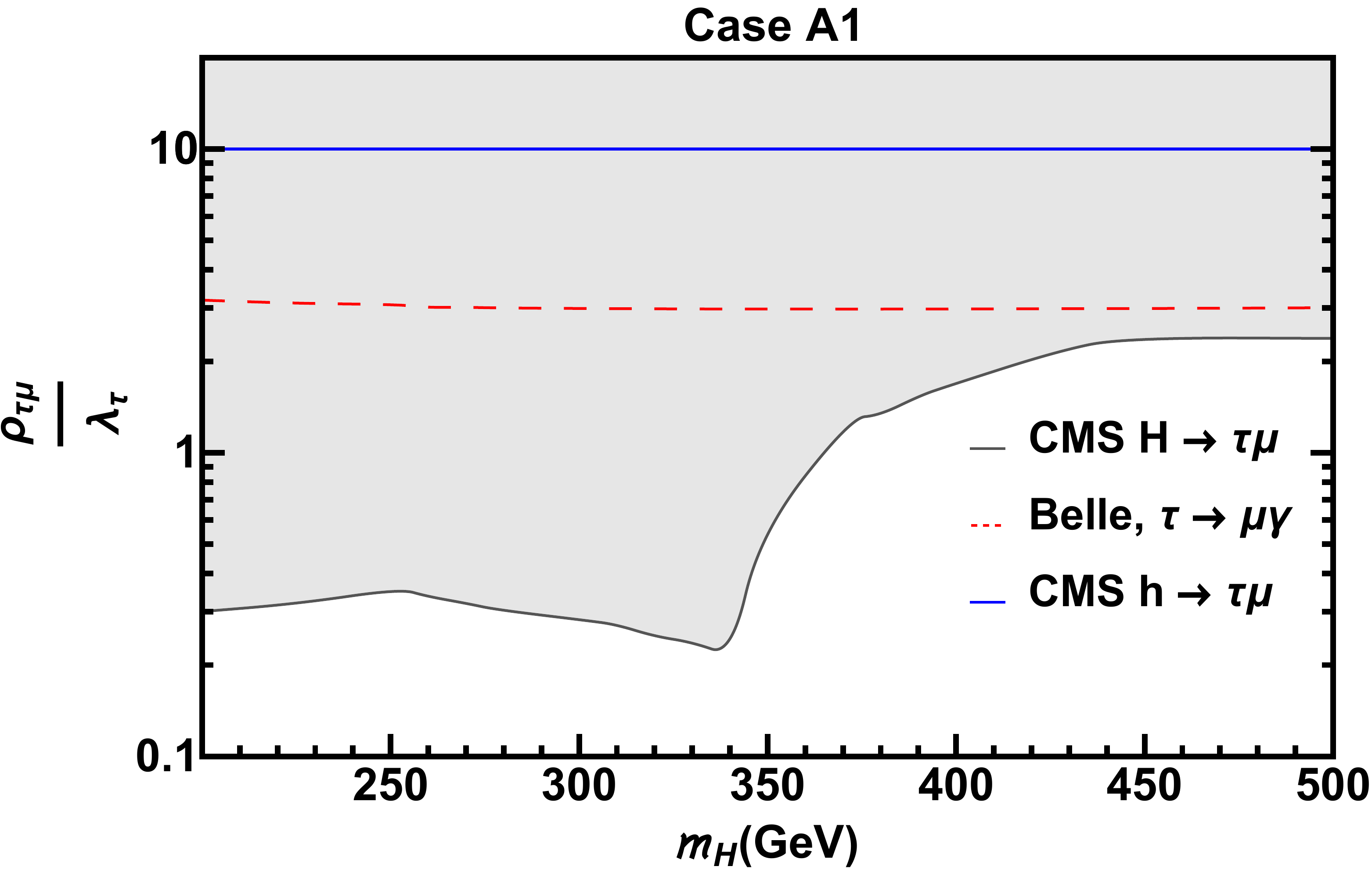}
\includegraphics[scale=0.15]{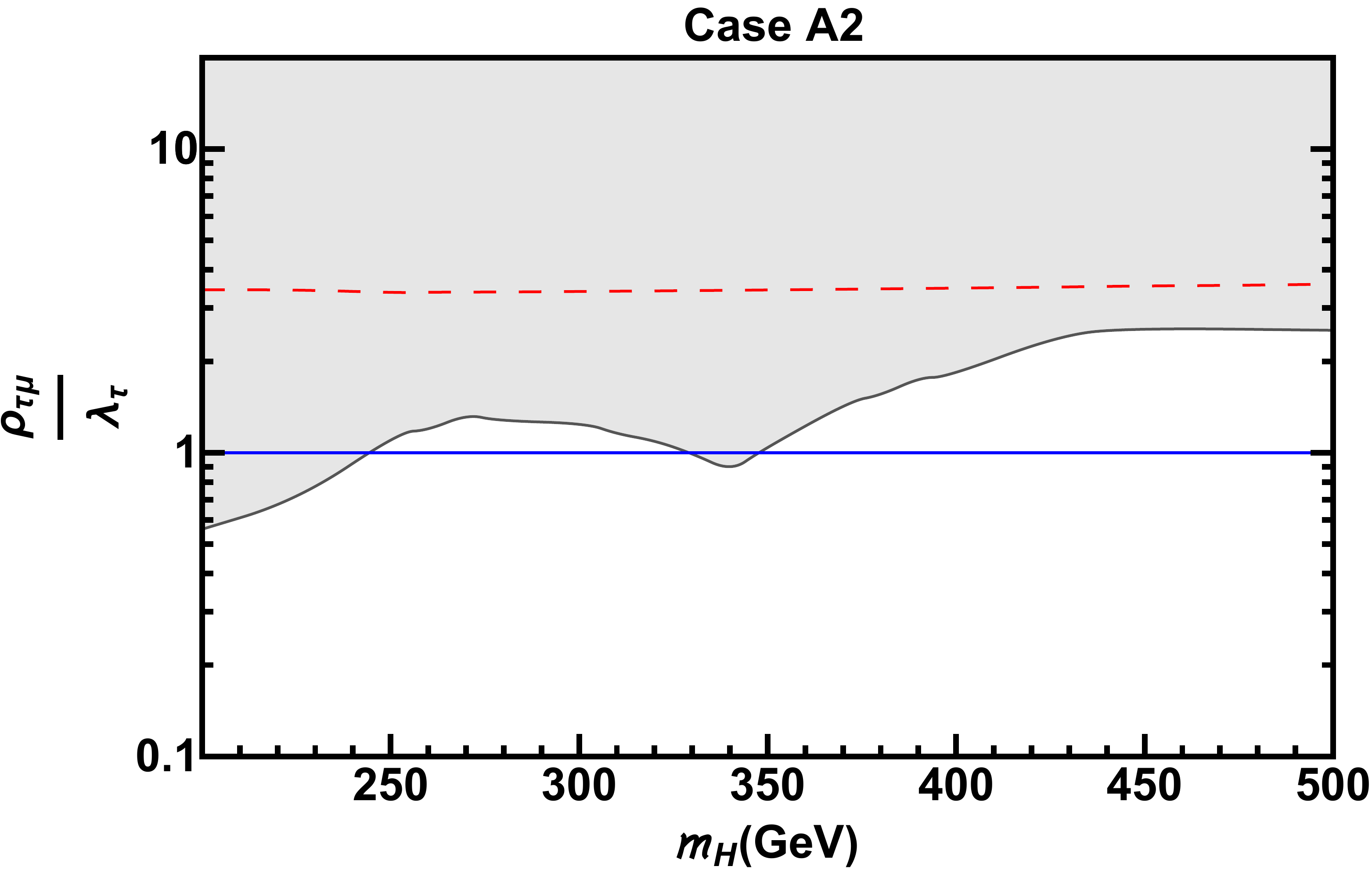} \\
\includegraphics[scale=0.15]{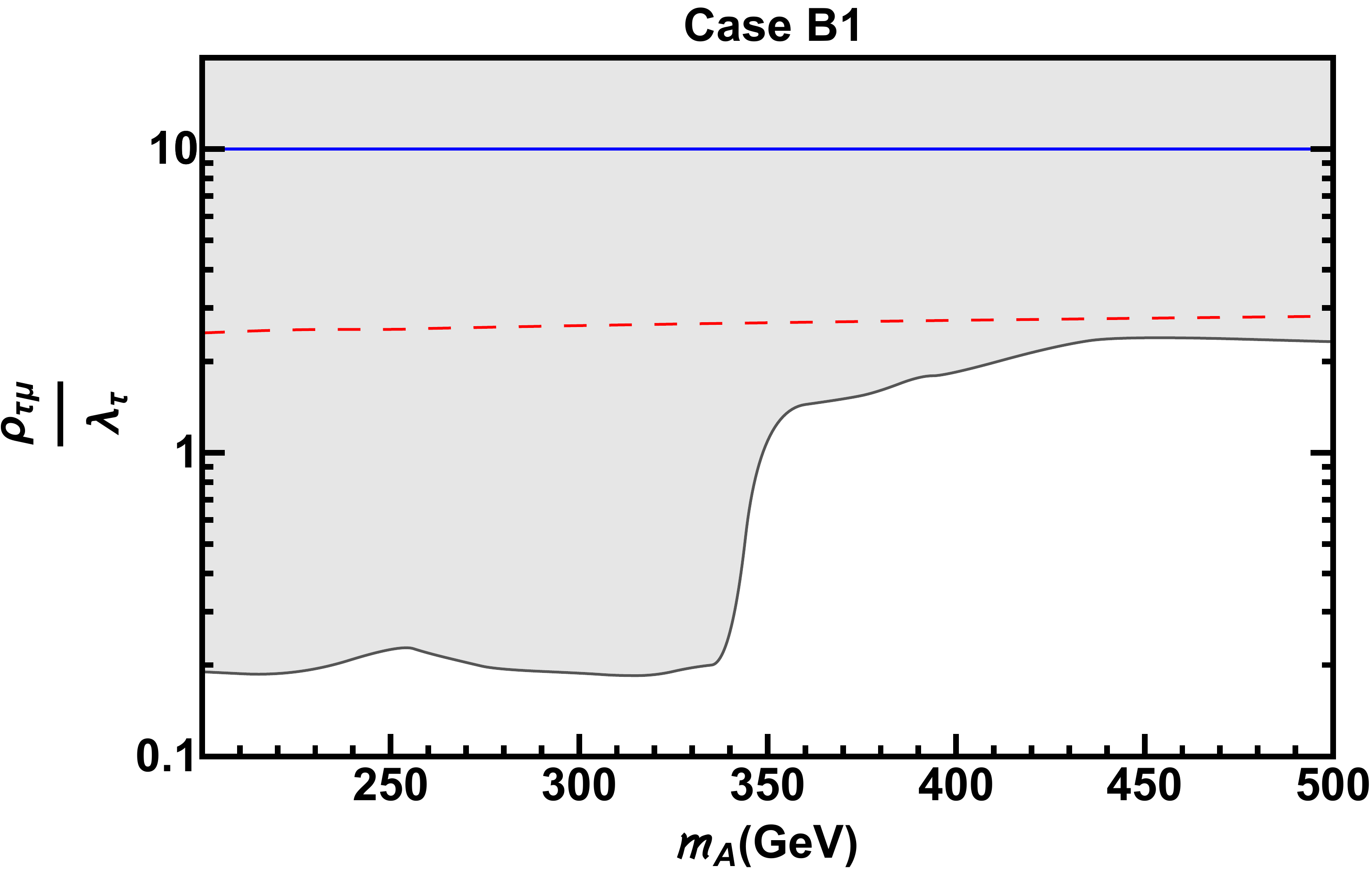}
\includegraphics[scale=0.15]{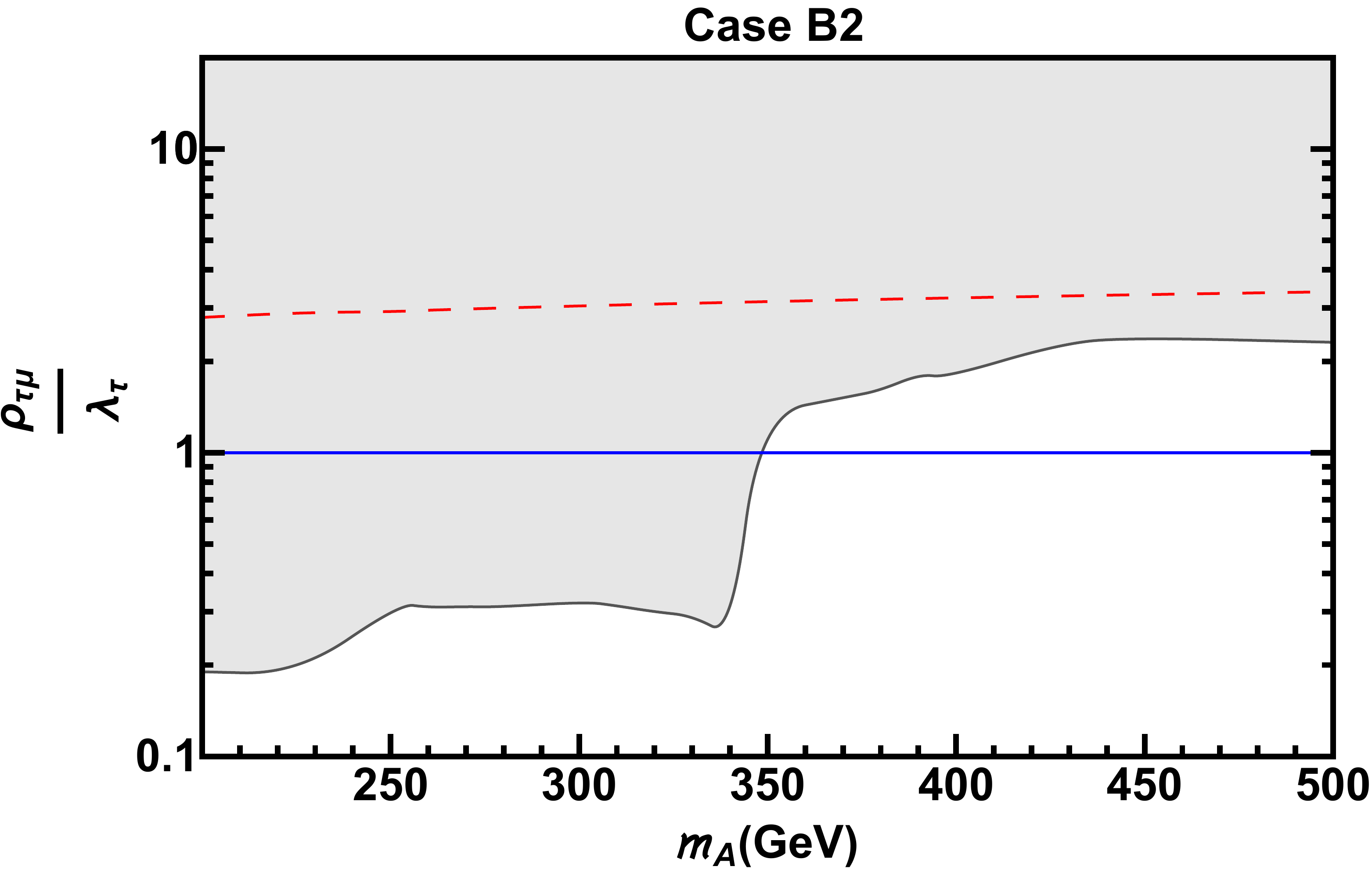} \\
\caption{
  Limits on $\rho_{\tau\mu}$ from $H\to \tau \mu$ (gray-shaded),
  $h \to \tau\mu$ (blue-solid) and $\mathcal{B}(\tau\to \mu
  \gamma)$ (red-dashed).
}
\label{fig:lim_rho_lm}
\end{figure*}

Using the branching fractions from Fig.~\ref{BF}, 
we estimate the limits on 
$\rho_{\tau\mu}$ ($\rho_{\tau\mu}
 = \rho_{\mu\tau}$) from CMS, and find it 
to be more stringent than from Belle,
where $\rho_{\tau\mu}$ and $\rho_{\mu\tau}$ 
also receive constraints from flavor physics, 
the most relevant one from $\tau \to \mu \gamma$. 
Belle recently measured~\cite{Belle:2021ysv}
$\mathcal{B}(\tau\to\mu\gamma) < 4.2 \times 10^{-9}$ 
at 90 \% C.L., improving slightly over the 
BaBar limit of $\mathcal{B}(\tau\to\mu\gamma)
 < 4.4 \times 10^{-9}$~\cite{BaBar:2009hkt} 
at 90\% C.L. The branching fraction of 
$\tau \to \mu \gamma$ is~\cite{Omura:2015xcg},
\begin{equation}
    \mathcal{B}(\tau \to \mu \gamma) = \frac{48 \pi^3 \alpha}{G_F^2} (|A_L|^2 + |A_R|^2)\mathcal{B}(\tau \to \mu\nu_{\tau}\bar{\nu}_{\mu}),
\end{equation}
where {$\mathcal{B}(\tau \to \mu\nu_{\tau}\bar{\nu}_{\mu})
 = 17.39\%$}~\cite{ParticleDataGroup:2020ssz}, 
and $A_{L(R)}$ are the amplitudes based on different 
chiral structures coming from one- and two-loop diagrams. 
We include one-loop effects from all $A$, $H$ and $H^+$, 
and likewise for Barr-Zee type two-loop contributions. 
Bounds on $\rho_{\tau\mu}$ from Belle is given in 
Fig.~\ref{fig:lim_rho_lm} red (dashed) for all four cases. 
We find the Belle bound from $\tau \to \mu \gamma$ 
is {\it weaker} than the CMS bound from $H \to \tau\mu$, 
and for all cases $\rho_{\tau\mu}$ can be lower 
than $\lambda_{\tau}$ below 2$m_t$ threshold; 
the bounds for lighter $A$ are even more stringent 
than lighter $H$ due to higher production cross section
for the same values of mass.  
The $\mathcal{B}(\tau\to \mu\gamma)$ bounds show 
little mass dependence, largely because 
of our $\rho_{tt}$ from Eq.~(\ref{eq:rho_tt}), 
increasing $\rho_{tt}$ compensates the suppression
from heavier scalar mass of the two-loop diagram.
We keep $\rho_{\tau\tau} = \lambda_{\tau} \sim 0.01$.

\begin{figure*}[t]
\includegraphics[scale=0.15]{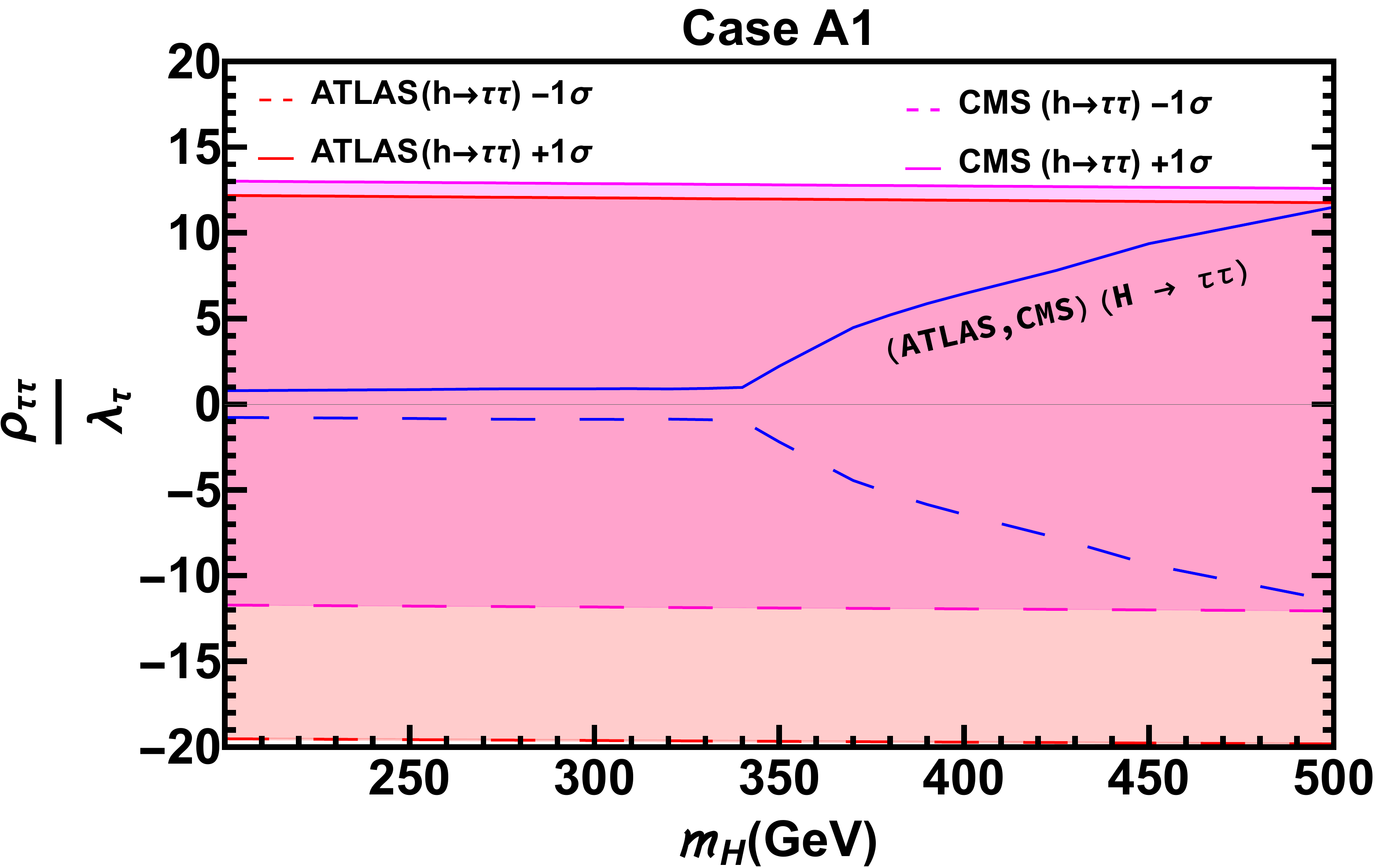}
\includegraphics[scale=0.15]{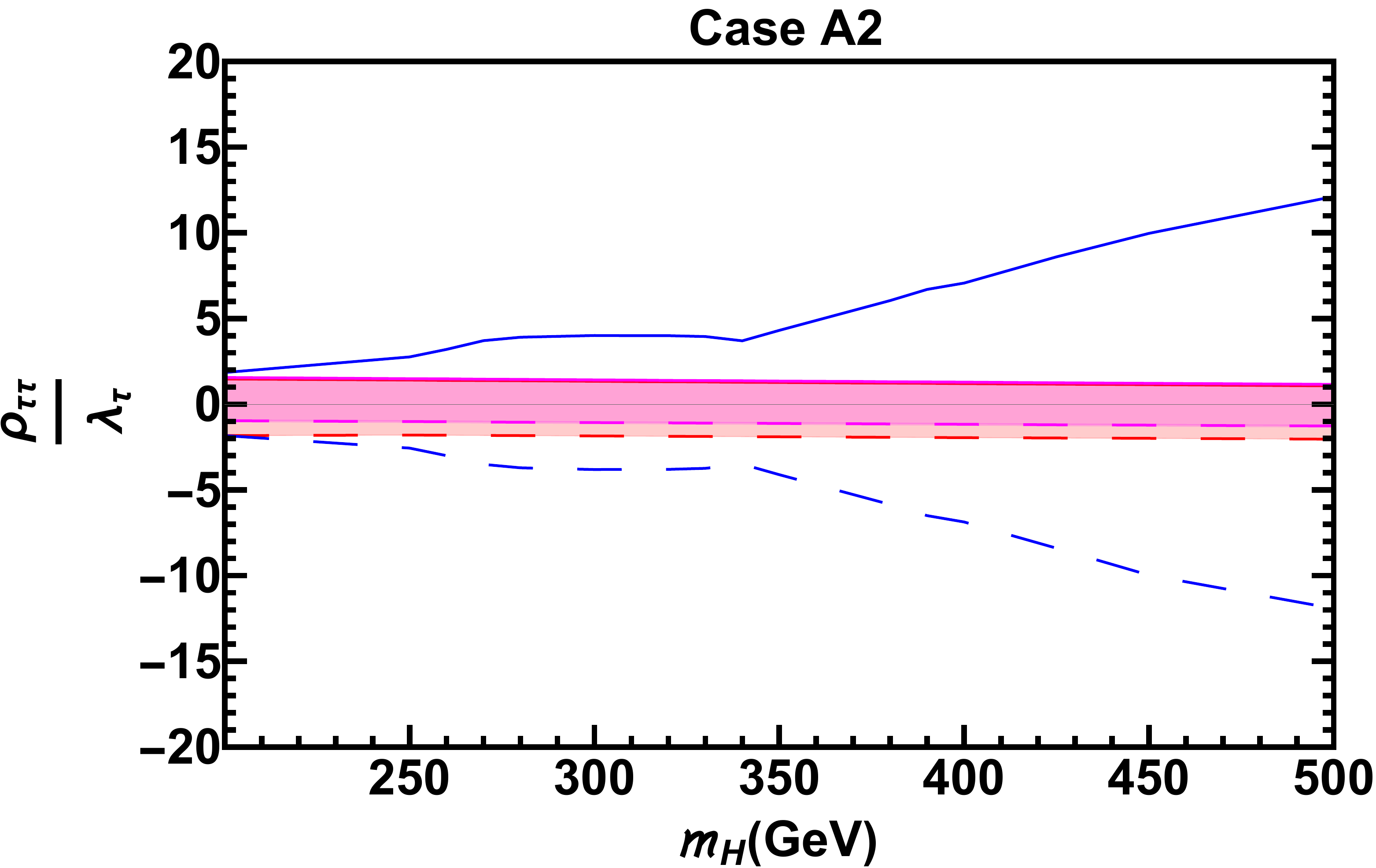} \\
\includegraphics[scale=0.15]{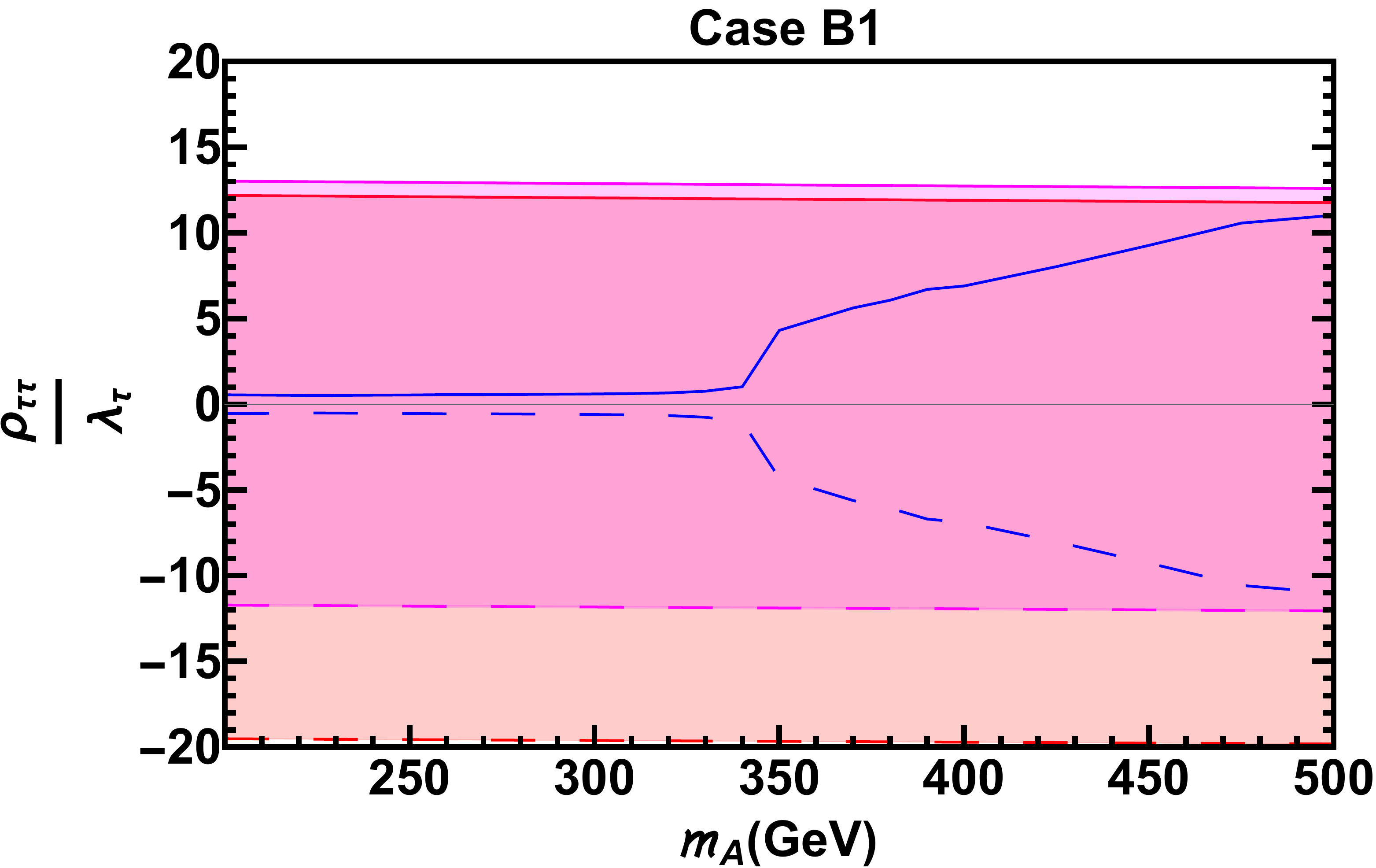}
\includegraphics[scale=0.15]{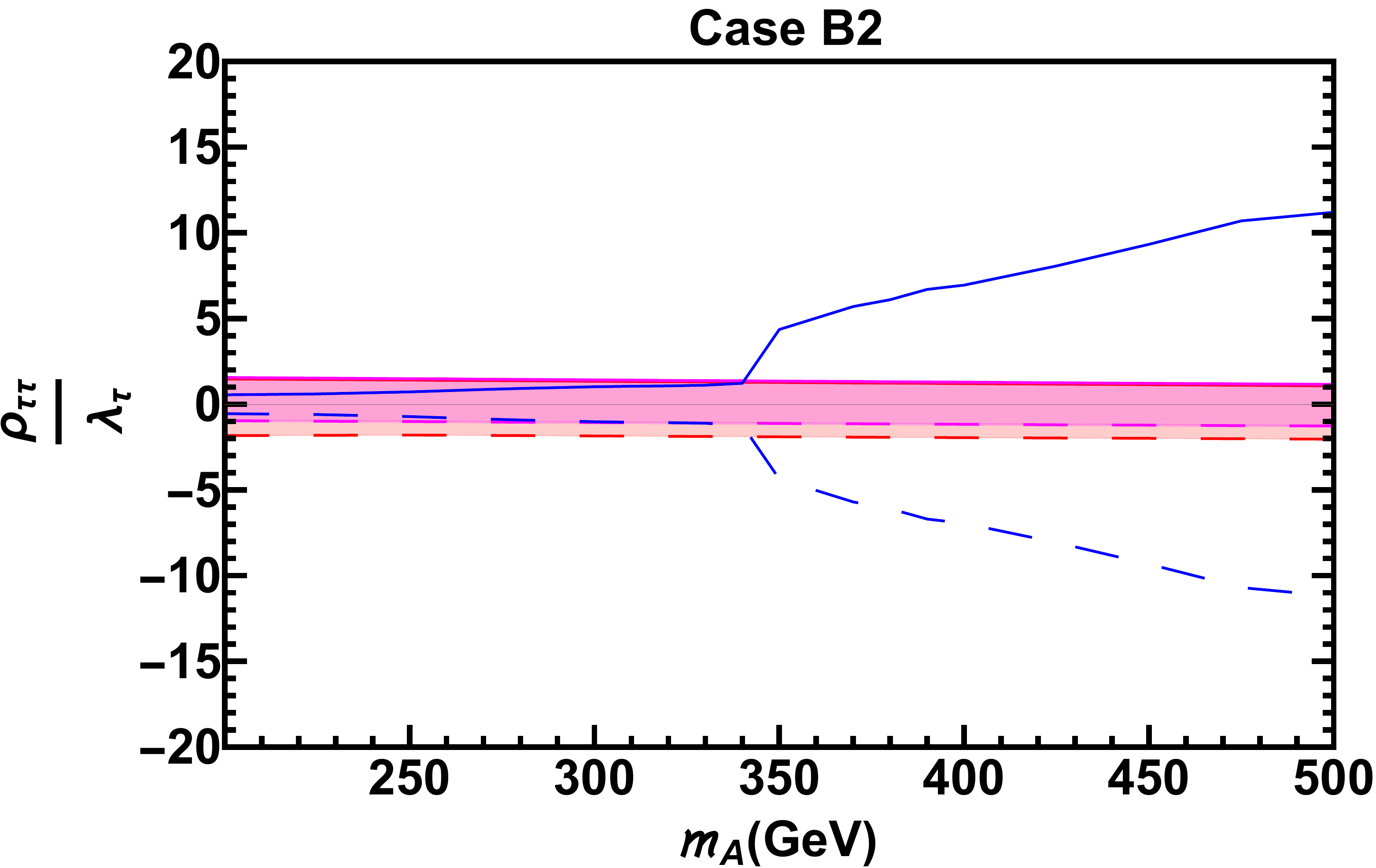} 
\caption{
Limits on $\rho_{\tau\tau}$ from CMS~\cite{CMS:2017zyp} (salmon region)
  and ATLAS~\cite{ATLAS:2018ynr} (pink region) measurements of
  $\mathcal{B} (h \to \tau\tau)$ to within one standard deviation.
  Also shown are the bounds from
  CMS~\cite{CMS:HTATA} and ATLAS~\cite{ATLAS:HTATA}
  searches for $p p \to H \to \tau \tau +X$ (blue lines). 
}
\label{fig:lim_rho_ll}
\end{figure*}

For $\rm c_{\gamma} \neq 0$, slightly away from 
the alignment limit, $pp \to h \to \tau\tau$ also 
puts constraints on $\rho_{\tau\tau}$ along with 
$p p \to H \to \tau\tau$ direct search by ATLAS and CMS.  
From Fig.~\ref{lhc_limit}, we select min(CMS, ATLAS) 
and use the $\rho_{tt}$ ansatz of Eq.~(\ref{eq:rho_tt}) 
to estimate the bounds on $\rho_{\tau\tau}$ for 
the four cases of Eq.~(\ref{eq:cases}). 
Our results are presented in Fig.~\ref{fig:lim_rho_ll}, 
where we keep $\rho_{\tau\mu} = \lambda_{\tau}$.

\begin{figure*}[t]
\includegraphics[scale=0.15]{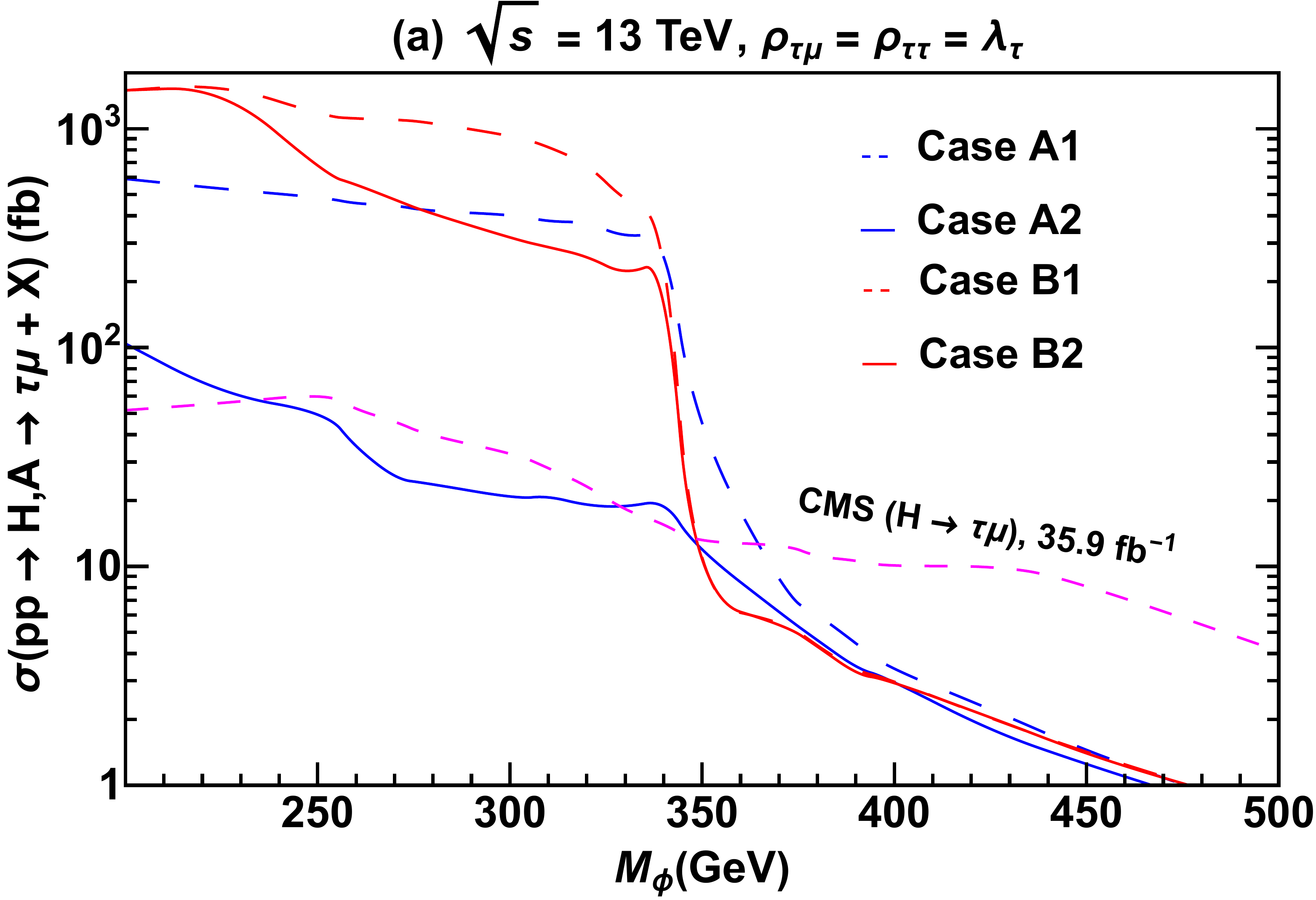}
\includegraphics[scale=0.15]{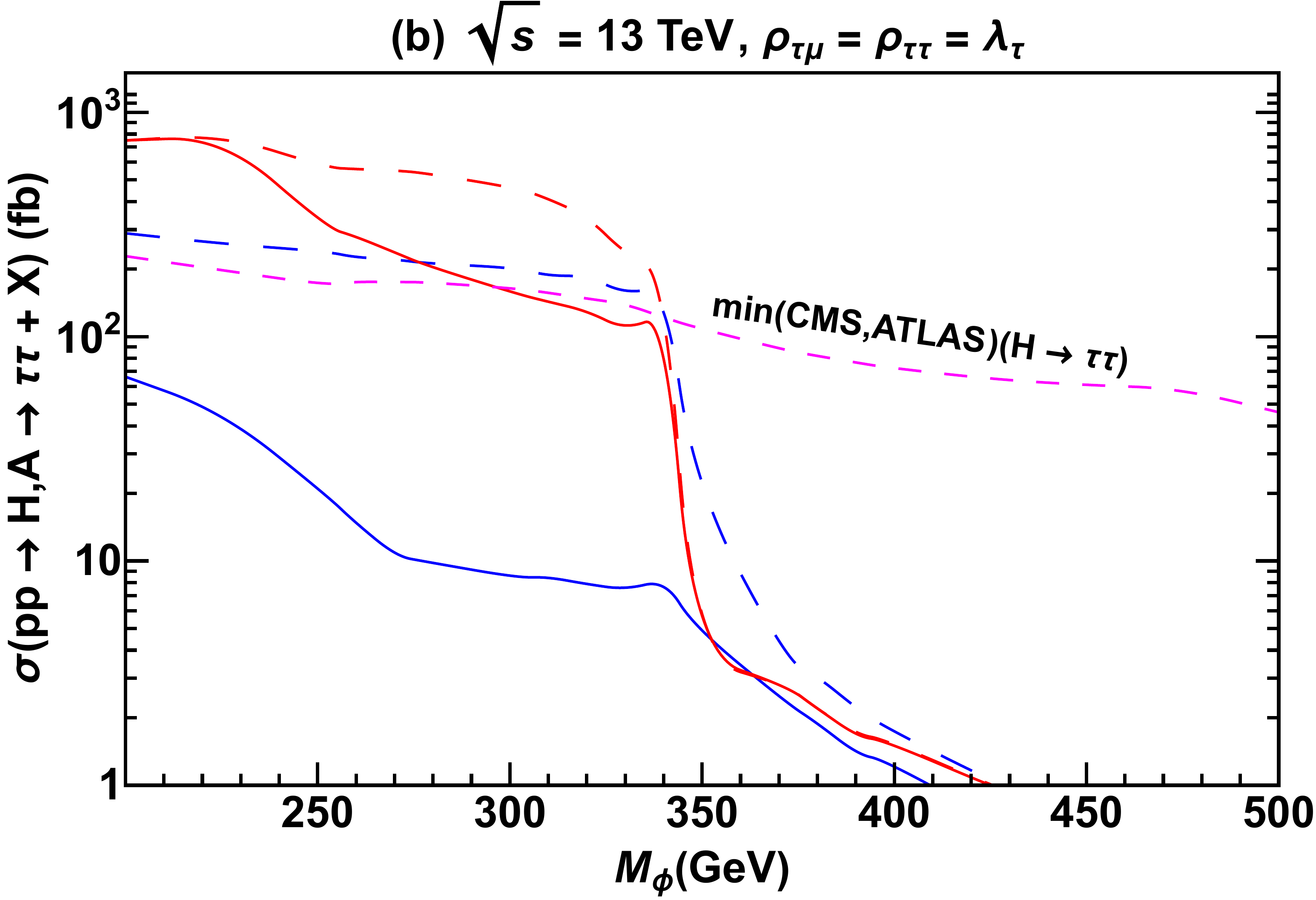}
\caption{
 Production cross sections for
 (a) $pp \to H,A \to \tau \mu + X$, and 
 (b) $\tau\tau + X$ for $\sqrt{s}$ = 13 TeV. 
 The CMS2019 limits on $ \tau \mu$ cross section
 and min(CMS,ATLAS) limits on $\tau\tau$ are also shown. 
}
\label{phitamu_fig}
\end{figure*}

The $\rho_{\tau\mu}$ and $\rho_{\tau\tau}$ bounds 
are correlated, but since neither $\tau\mu$ 
nor $\tau\tau$ are the dominant decay mode, 
a closer look at the limits of 
Figs.~\ref{fig:lim_rho_lm} and \ref{fig:lim_rho_ll}, 
we find that they do not deviate much above $\mathcal{O}(\lambda_{\tau})$. 
We have checked that increasing one 
does not significantly affect the limits 
on the other from CMS and ATLAS searches. 
For $\tau\to \mu \gamma$, $\rho_{\tau\tau}$ 
only comes at one-loop level which is chiral-suppressed, 
hence Belle limits on $\rho_{\tau\mu}$ have no effect. 
We present the production cross sections for 
$pp\to H,A \to \tau\mu$ and $pp \to H,A \to \tau\tau$ 
in Figs.~\ref{phitamu_fig}(a) and \ref{phitamu_fig}(b), respectively, 
using the $\rho_{tt}$ ansatz of Eq.~(\ref{eq:rho_tt}) 
and branching fractions from Fig.~\ref{BF}.

\section{\boldmath Collider prospects for 
 \large{$H,A \to \tau\mu$}}
%\begin{itemize}
%\item Signal Considered  $pp\to \phi \to \tau\mu \to e \mu + \tau_{j}\mu + X$

In this section we demonstrate our approach towards 
searching for the $H,A \to \tau\mu$ channel at LHC. 
For $\tau$ decay, we include $\tau \to e \nu_{e} \nu_{\tau}$ 
and $\tau \to j_{\tau} \nu_{\tau}$ decay modes, 
where $j_{\tau} = \pi, \rho, a_1$. 
We divide our collider study into two parts,
 (a) fully leptonic channel, $pp \to H,A\to\tau\mu \to e\mu + \slashed{E}_T + X$, and
 (b) semileptonic channel, $pp \to H,A \to \tau \mu \to j_{\tau}\mu + \slashed{E}_T + X$. 
We consider all four cases of Eq.~(\ref{eq:cases}). 
For simplicity we keep 
$\rho_{\tau\mu} = \rho_{\tau\tau} = \lambda_{\tau}$, 
but for estimating statistical significance, 
we follow the limits derived in previous section.

\vskip0.175cm
\noindent{\bf Analysis procedure and event generation.}
Two types of collider studies are performed:
(a) parton level (PL) without hadronization or detector effects;
 (b) event level (EL) with hadronization using PYTHIA\,8.2~\cite{pythia8}
  and detector effects simulated by DELPHES\,3.5~\cite{delphes}. 
For the parton level analysis, we use our code for phase-space integration 
using the VEGAS algorithm~\cite{vegas}. 
For the signal, we use analytic expressions 
from~\cite{Kao:1993du} to calculate $gg \to H,A$ 
production at tree level, then use HIGLU~\cite{higlu} 
to estimate higher-order corrections. 
We use CT14LO~\cite{Dulat:2015mca} parton 
distribution functions to calculate leading order (LO) processes.

For the backgrounds, we use MadGraph5~\cite{Alwall:2011uj}
and HELAS~\cite{Hagiwara:2008jb} libraries to extract 
matrix elements for all possible Feynman diagrams at LO 
and scale them using $K$-factors. 
We apply minimal smearing of lepton and jet momenta 
following ATLAS~\cite{atlas_res} and CMS~\cite{CMS:2016lmd} 
specifications. For simplicity, we keep 
the smearing for $e$ and $\mu$ the same:
\begin{equation}
\frac{\Delta E}{E} = \frac{0.6}{\sqrt{E(\rm GeV)}} \oplus 0.03 \ (\rm jets), \quad
%\end{equation}  
%
%\begin{equation}
\frac{\Delta E}{E} = \frac{0.25}{\sqrt{E(\rm GeV)}} \oplus 0.01 \ (\rm leptons).
\end{equation} 
We use collinear approximation~\cite{coll_app} for $\tau$ decay at the
parton level.

For the event level analysis, we first generate parton level events
with Madgraph, then pass it to PYTHIA\,8.2 and then DELPHES\,3.5. 
We use the anti-$k_T$ algorithm~\cite{antikt} for 
jet clustering and keep all parameters at default values, 
as described in the DELPHES card for the CMS detector. 
The decays of $\tau$ leptons are modeled using TAUOLA~\cite{Jadach:1990mz}.

\vskip0.175cm
\noindent{\bf Fully leptonic channel and backgrounds.}
With $\tau \to e \nu_e \nu_{\tau}$, our signal is 
$pp \to H,A \to \tau\mu \to e \mu + \slashed{E}_T + X$. 
So in the final state we have two opposite sign, different 
flavor leptons along with missing transverse energy. 
Important backgrounds come from $W^+W^-$, $t\bar{t}$, 
$tW^{\pm}$ and $Z,\gamma^* \to \tau \tau$. 
We use TOP++ to estimate the $K$-factor 
for $t\bar{t}$ background~\cite{toppp}, 
and MCFM 8.0~\cite{mcfm} to estimate the 
NLO corrections to remaining backgrounds. 
Since $\mu$ comes directly from Higgs decay, 
it is quite energetic, so we select events 
with $p_{T}(\mu) > $ 60 GeV and $|\eta(\mu)| <$ 2.4. 
For electron, we select events with 
$p_{T}(e) > $ 10 GeV, $|\eta(e)| < $ 2.4. 
In addition, we veto all events with extra jets 
and require $\slashed{E}_T > $ 20 GeV.

\begin{table}[t]
    \centering
    \begin{tabular}{cll} \hline
    Variables \ & \ Leptonic & \ Semileptonic  \\ \hline
    $p_{T}(e)$  & \ $>$ 10 GeV & \ \ NA \\
    $p_{T}(\mu)$ & \ $>$ 60 GeV & \ \ $>$ 60 GeV \\
    $p_T(j_{\tau})$ & \ NA & \ \ $>$ 30 GeV \\
    $|\eta(e)|$ & \ $<$ 2.4 & \ \ NA \\
    $|\eta(\mu)|$ & \ $<$ 2.4 & \ \ $<$ 2.4 \\
    $\slashed{E}_T$ & \ $>$ 20 GeV & $>$ 20 GeV \\
    $M_T(e,\slashed{E}_{T})$ & \ $<$ 100 GeV & \ \ NA \\
    $M_T(\mu, \slashed{E}_T)$ & \ $>$ 100 GeV & \ \ $>$ 100 GeV \\
    $M_T(j_{\tau},\slashed{E}_{T})$ & \ NA & \ \ $<$ 105 GeV \\
    $|M_{\tau\mu} - m_H|$ & \ $< 0.2\, m_H$ &  \ \ $< 0.2\, m_H$ \\
    $|\Delta \phi(e,\slashed{E}_T)|$ & \ $<$ 1.0 & \ \ NA \\ 
    $\Delta R(j_{\tau},\mu)$& \  NA  & \ \ $>$ 0.4 \\ 
    $N_j$ & \ NA & \ \ 0 (PL)  \\ \hline 
    \end{tabular}
\caption{
  Cuts applied for leptonic and semileptonic channels
  in $H, A\to \tau \mu$ study.}
\label{tab:cutstm}
\end{table}

   For selected events, we reconstruct
  the transverse mass~\cite{Barger:1987nn} of a lepton ($\ell$) 
  and missing transverse energy ($\slashed{E}_T$),
  $M_{T}(\ell, \slashed{E}_T), \ell = e$ or $\mu$, with
\begin{equation}
M^2_{T}(\ell,\slashed{E}_T) 
 = (p_{T}(\ell) + \slashed{E}_T)^2
   - (\vec{p}_T(\ell) + \vec{\slashed{E}}_T)^2,
\label{transmass}
\end{equation}
 where $p_{T}(\ell)$ is the transverse momentum of electron or muon,
while $M_{\tau\mu}$ is the reconstructed invariant mass of $\tau$ and $\mu$
with a pronounced peak near $m_\phi, \phi = H$ or $A$,
using collinear approximation~\cite{Hagiwara:1989fn}. 
In the collinear approximation, the $\tau$ coming 
from $H, A$ decay is highly boosted, hence we assume 
that its decay products are also boosted in 
the same direction. Under the collinear approximation, 
we can write,
\begin{equation}
   p_{\rm vis} = x\, p_{\tau}\, , \quad {\rm and} \quad 
   p_{\rm \nu} = (1 - x)\, p_{\tau},
\label{eq:coll_app}
\end{equation}
where $p_{vis}$ is the four-momentum of 
the visible particle(s) from $\tau$ decay, 
$p_{\rm \nu}$ is the total four momentum of 
all neutrinos from $\tau$ decay, and $x$ is 
the fraction of $\tau$ momentum carried by $p_{\rm vis}$. 
We know the four-momentum of the visible particles
and $\slashed{E}_{T}$ coming from the $\nu$'s. 
After some algebra, we find
\begin{equation}
%  \tcr{x = \frac{p_{T}(\rm vis)}{p_{T}(\rm vis) + \slashed{E}_T}.(?)}\, \
   x = \frac{p_{T}(\rm vis)}
    { \sqrt{ ( p_x({\rm vis})+\slashed{E}_x )^2
            +( p_y(\rm vis)+\slashed{E}_y )^2 } }  \, .   
\label{eq:mom_fac}
\end{equation}
Note that this assumption will only give 
reasonable results if the $\tau$ lepton is 
the only source of missing transverse energy, 
and that it is highly boosted. 
We select events that satisfy
 $M_{T}(e, \slashed{E}_T) < $ 100 GeV and 
 $M_{T}(\mu, \slashed{E}_T) > $  100 GeV~\cite{CMS:2019pex}.

\begin{table}[t]
    \centering
    \begin{tabular}{clllllll}
%    \multicolumn{5}{c} \\ \hline
    \multicolumn{6}{c}{{\it Parton Level} \,($pp \to H,A \to \tau \mu$)} \\ \hline
    Backgrounds/$m_H$ & \ \ 200 GeV \ \ & 250 GeV & \ 300 GeV & \ 350 GeV & \ 400 GeV & \ 450 GeV & \ 500 GeV \\ \hline \hline
    $W^+W^-$ & \ 9.73 fb & 11.6 fb &  11.0 fb & \ 9.6 fb & \ 7.8 fb & \ 6.4 fb & \ 5.1 fb  \\
     $Z,\gamma \to \tau\tau$ & \  6.36 fb & \ \ 4.2 fb & \ \ 2.7 fb & \ 1.8 fb & \ 1.2 fb & \ 0.8 fb & \ 0.6 fb \\

    $tW^{\pm}$ & \ \ \ 3.1 fb & \ \ 4.2 fb & \ \ 4.0 fb & \ 3.7 fb & \ 2.9 fb & \ 2.4 fb & \ 1.9 fb \\
       $t\bar{t}$ & \ \ \ 3.1 fb & \ \ 3.7 fb & \ \ 3.3 fb & \ 2.7 fb & \ 2.3 fb & \ 1.8 fb & \ 1.4 fb \\
    \hline
    Total & \ 22.3 fb & \ 23.7 fb &  \ 21.1 fb & 17.8 fb & 14.5 fb & 11.4 fb & \ 9.0 fb \\ \hline \hline
    \multicolumn{6}{c}{{\it Event Level} ($pp \to H,A \to \tau \mu$)} \\ \hline
    $W^+W^-$  & \ 4.87 fb & \ 5.8 fb & \ 5.4 fb & \ 4.5 fb & \ 3.4 fb &\ 2.9 fb &\ 2.5 fb  \\
    $Z,\gamma \to \tau\tau$  & \ 2.94 fb & \ 1.9 fb & \ 1.3 fb & \ 0.7 fb & \ 0.3 fb & \ 0.2 fb & \ 0.1 fb \\
    $tW^{\pm}$ & \ 1.18 fb& \ 1.7 fb & \ 1.8 fb &  \ 1.5 fb & \ 1.3 fb & \ 1.0 fb & \ 0.8 fb \\
    $t\bar{t}$ & \ 0.83 fb & \ 1 \ \, fb & \ 0.9 fb & \ 0.7 fb & \ 0.6 fb & \ 0.5 fb & \ 0.4 fb \\
    \hline
    Total  & \ 9.82 fb & 10.4 fb & \ 9.3 fb & \ 7.5 fb & \ 5.9 fb & \ 4.4 fb & \ 3.7 fb  \\ \hline
    \end{tabular}
\caption{
  Background cross sections for $e \mu$ final state 
  %after all cuts
  at parton and event levels.}
 \label{crossx_lep_tm_bkgs}
 \end{table}
% Mention k-factors    

\begin{figure*}[t]
\centering
\includegraphics[scale=0.15]{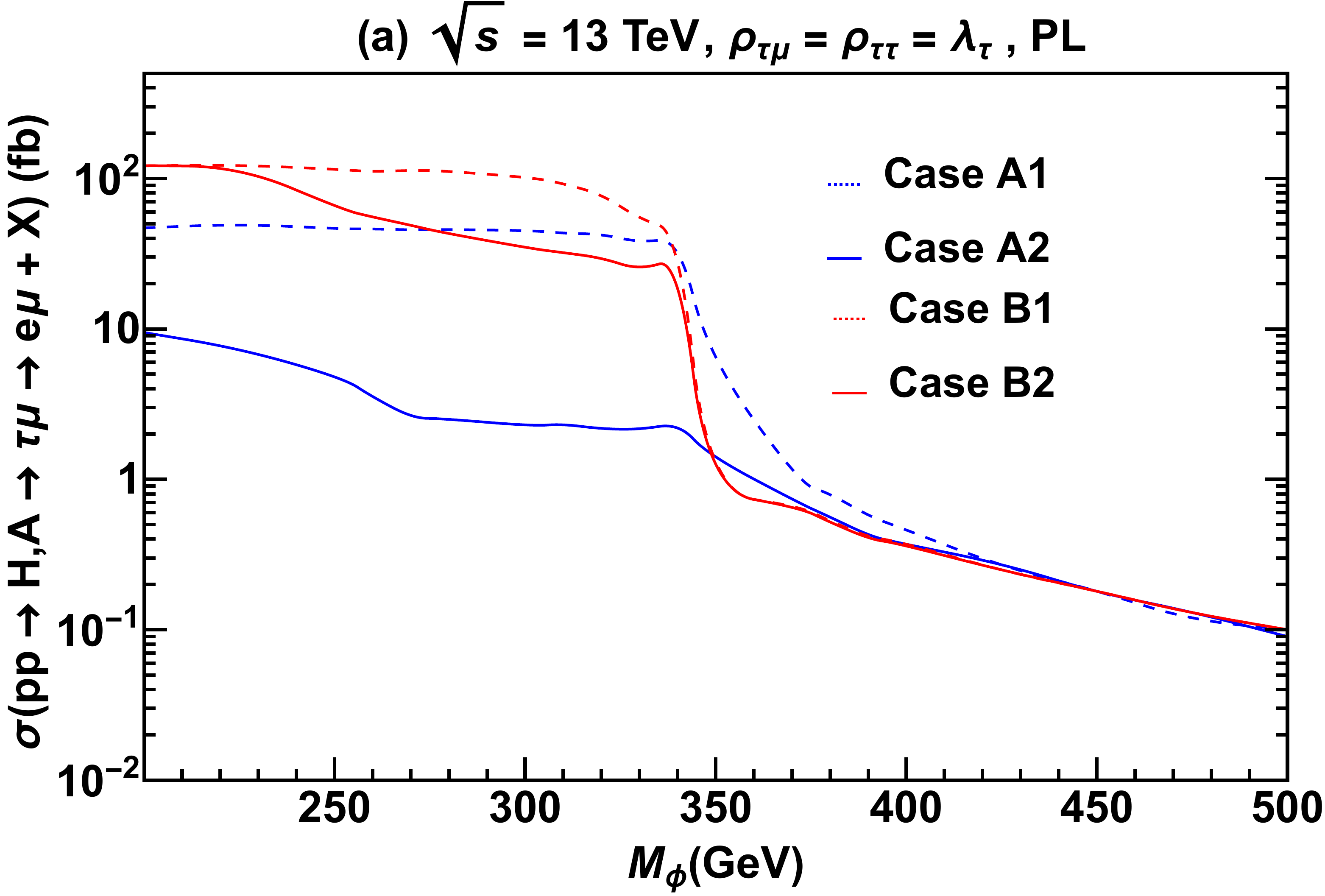}  
\includegraphics[scale=0.15]{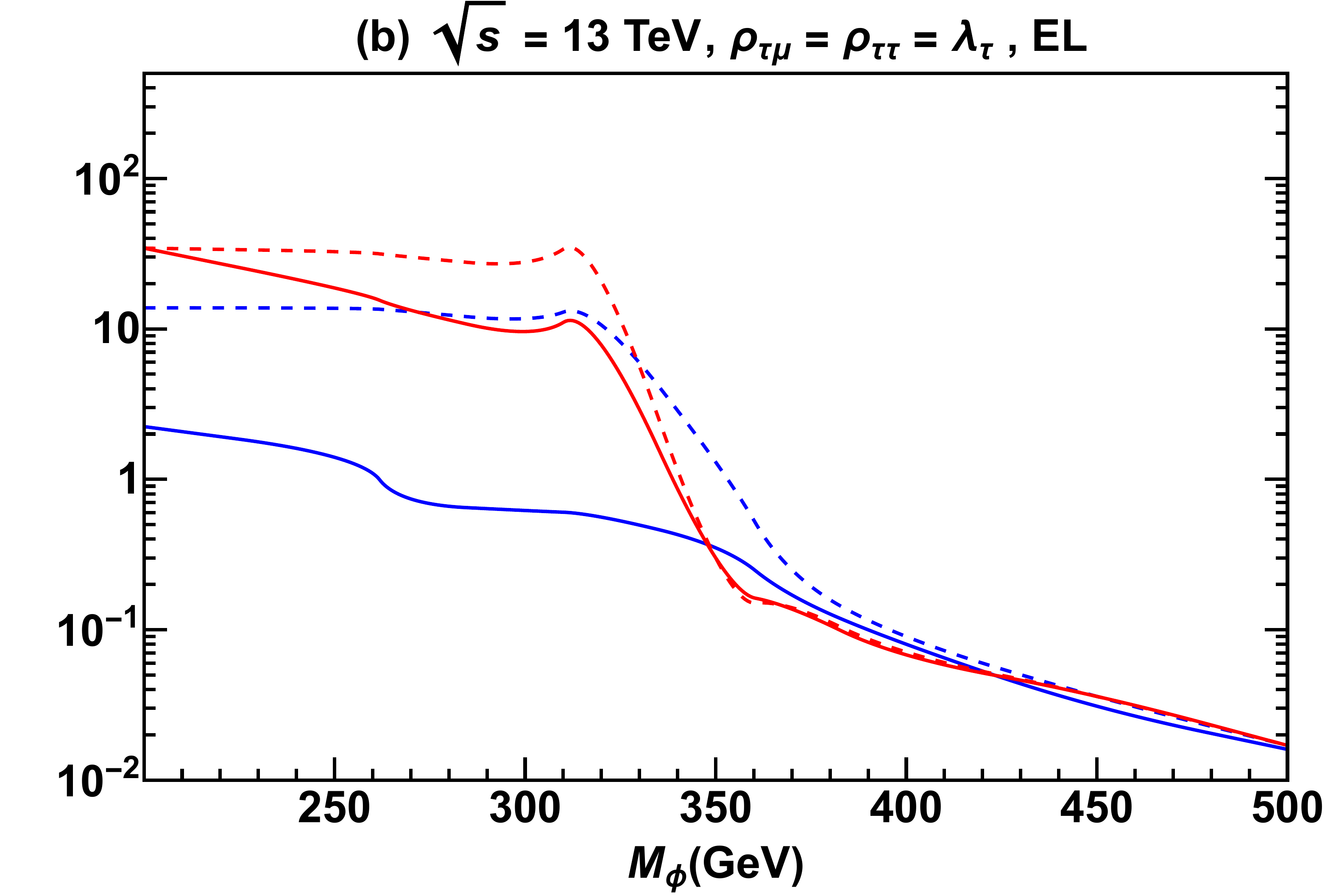}\\
\includegraphics[scale=0.15]{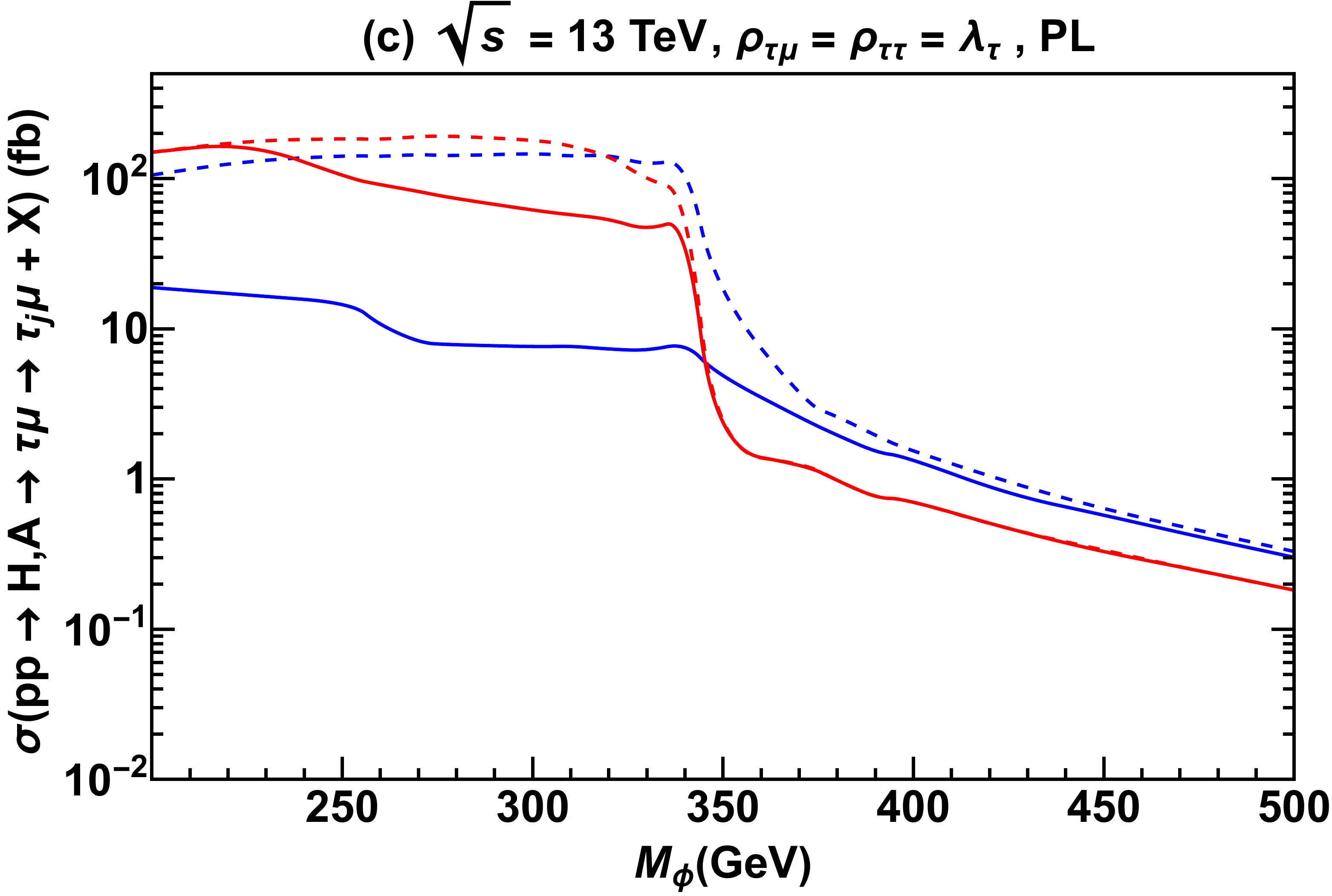}
\includegraphics[scale=0.15]{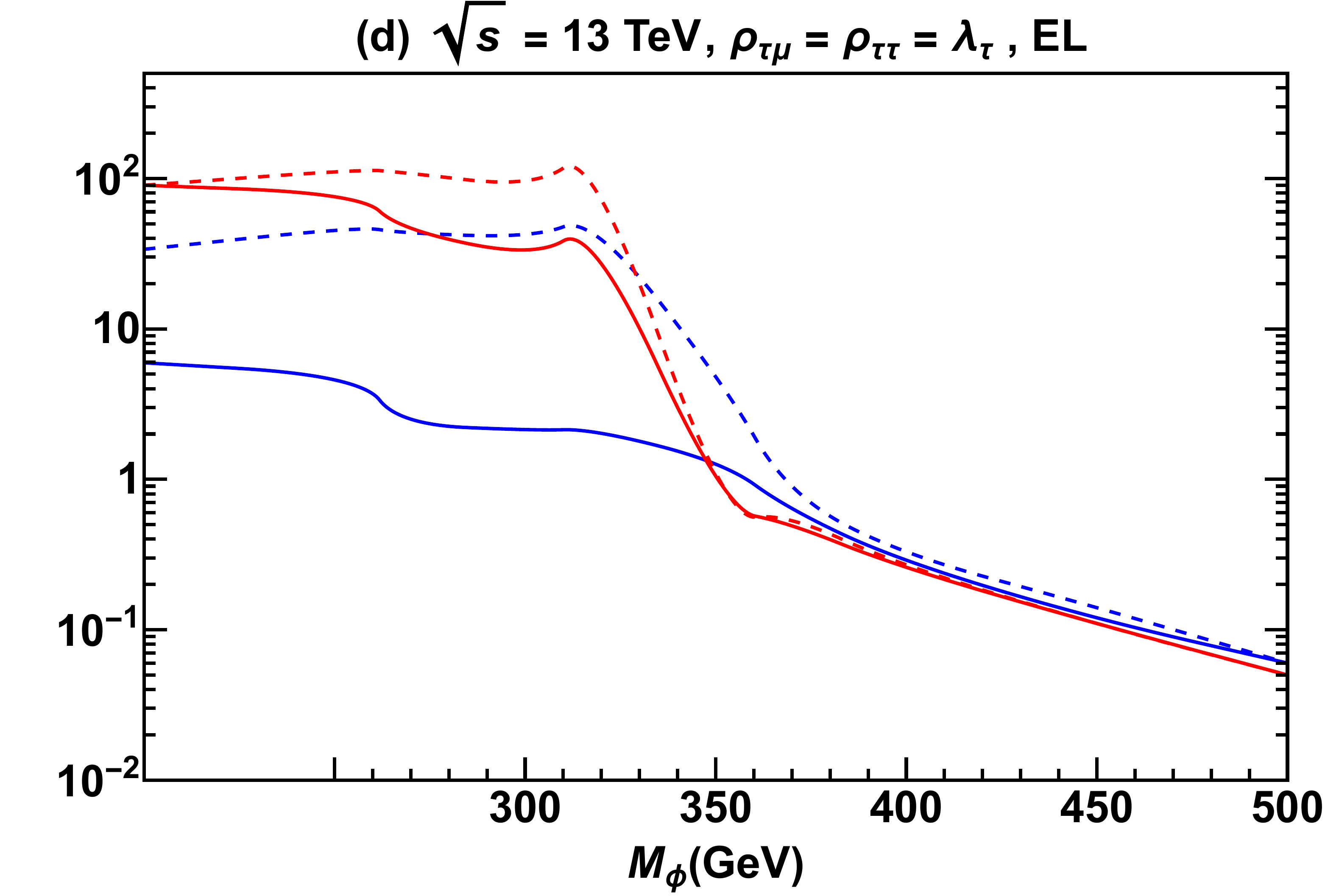}
\caption{
  Cross sections after all cuts at parton (a, c),
  and event (b, d) levels for $p p \to H,A \to \tau \mu$.
  Upper (lower) panels are for the (semi-)leptonic channel.
} 
\label{fig:crossxtm}
\end{figure*}

 We define a moving mass window of 
$|M_{\tau\mu} - m_{H,A}| < 0.2 \times m_{H,A}$ 
  to estimate the irreducible background from SM processes
  for a particular $m_H$. 
All the cuts discussed here are summarized in Table~\ref{tab:cutstm}. 
Cross sections of all backgrounds for PL and EL are 
presented in Table~\ref{crossx_lep_tm_bkgs}. 
The signal cross sections are given 
in Figs.~\ref{fig:crossxtm}(a) and (b). It is important to note that two b-veto plays a vital role in 
suppressing $t\bar{t}$ background. In the CMS~\cite{CMS:2019pex} study, their $t\bar{t}$ enriched control region requires 
at least 1-jet to be tagged as a b-jet. We have checked that when we select events that contain at least 1 b-jet the 
$t\bar{t}$ becomes the most dominant background almost 20 times the
contribution from $W^+W^-$.

\vskip0.175cm
\noindent{\bf Semileptonic channel and backgrounds.} 
When $\tau$ decays hadronically to $j_{\tau}$, 
the signal becomes $p p \to H,A \to \tau \mu
 \to j_{\tau} \mu + \slashed{E}_T + X$, 
giving us a final state with one jet tagged 
as a $\tau$-jet, one $\mu$ and missing transverse energy. 
Important backgrounds come from $W^{\pm}j$, $t\bar{t}$, 
$W^+W^-$, $Z,\gamma^* \to \tau \tau$ and $tW^{\pm}$. 
Event selection is similar to the leptonic channel.
Following CMS~\cite{CMS:2019pex}, we require 
$p_{T}(j_{\tau}) > $ 30 GeV and $|\eta(j_{\tau})| < 2.5$,
with muon selection the same as before. 
We again reconstruct the transverse masses 
$M_T(\mu, \slashed{E}_T)$ and $M_T(j_{\tau},\slashed{E}_T)$, 
then the collinear mass $M_{\tau\mu}$ 
using Eqs.~(\ref{eq:coll_app}) and (\ref{eq:mom_fac}) 
to reconstruct $\tau$ four momentum. 
Cuts on the reconstructed transverse masses are 
taken from CMS~\cite{CMS:2019pex} and summarized 
in Table~\ref{tab:cutstm}. 
Background cross sections after all cuts in Table~\ref{tab:cutstm} are applied for 
different Higgs masses are given in Table~\ref{crossx_slep_tm_bkgs}.
The signal cross sections as a function of Higgs mass 
are given in Figure~\ref{fig:crossxtm}(c) and \ref{fig:crossxtm}(d). 
We scale the parton level signal cross section 
and $Z,\,\gamma^* \to \tau \tau$ with 
$\epsilon_{j_{\tau}}$ = 0.7~\cite{Friis:2011zz}, 
and the mistag rates at 1/35 for 1-prong 
and 1/240 for 3-prong $\tau$ decays~\cite{ATLAS:2019uhp}. 
%
%In the next section we present a detailed analysis for the $H,A\to \tau \tau$ channel.

\begin{table}[h]
    \centering
    \begin{tabular}{clllllll}
%    \multicolumn{5}{c}{$pp \to H,A \to \tau \mu$} \\ \hline
    \multicolumn{6}{c}{{\it Parton Level} \,($pp \to H,A \to \tau \mu$)} \\ \hline
    Backgrounds/$m_H$ \ \ & 200 GeV \ \ & \ \ 250 GeV & \ \ 300 GeV & \ \ 350 GeV & \ \ 400 GeV & \ \ 450 GeV & \ \ 500 GeV \\ \hline \hline
    $W^\pm j$ & 210.42 fb  & \  192.7 fb &  \ 143.9 fb & \  98.6 fb & \ \ 66.4 fb & \ \ 45.2 fb & \ \ 31.3 fb  \\
     $t\bar{t}$ & \ \ \ 9.49 fb & \ \ \hspace{0.001mm} 13.5 fb & \  \hspace{0.2mm} 12.9 fb & \  10.4 fb & \ \ \ 7.8 fb & \ \ \ 5.6 fb & \ \  4.1 fb \\
    $W^+W^-$ &  \ \ \  3.22 fb & \ \ \ \  3.5 fb &  \ \  \hspace{1mm} 2.9 fb & \ \ \ 2.2 fb & \ \ \ 1.7 fb & \ \ \ 1.2 fb & \ \ 0.9 fb  \\
   $tW^{\pm}$ & \ \ \ 1.63 fb& \ \ \ \ 2.2 fb & \ \  \hspace{1mm} 2.2 fb & \ \ \ 1.9 fb & \ \ \ 1.4 fb & \ \ \ 1.1 fb & \ \ 0.9 fb \\
   
     $Z,\gamma \to \tau\tau$ & \ \ \ 4.14 fb& \ \ \ \ 2.1 fb & \ \ \hspace{1mm} 2.0 fb & \ \ \ 1.6 fb & \ \ \ 1.2 fb & \ \ \ 0.9 fb & \ \  0.7 fb \\ \hline
    Total & \ \ 228.9 fb&  \ \ 214.0 fb &  \hspace{1mm}163.9 fb & 114.6 fb & \ \ 78.5 fb & \ \ 54.0 fb & 37.8 fb \\ \hline \hline
 %   \multicolumn{5}{c}{\it Event Level} \\ \hline
 %   $W^\pm j$ &  197.3 fb &  155.1 fb & \ 90.2 fb & \ 70.5 fb & \ 47.9 fb & \ 36.7 fb  \\
 %    $t\bar{t}$ & \ \ 5.01 fb & \ \ 3.9 fb & \ 2.4 fb & \ 1.9 fb & \ 1.3 fb & \ 0.9 fb \\
 %   $W^+W^-$ & 0.97 fb & 0.73 fb & 0.43 fb & 0.25 fb & 0.24 fb & 0.23 fb  \\
   
 %   $tW^{\pm}$ & 0.58 fb & 0.44 fb & 0.23 fb & 0.15 fb & 0.15 fb  & 0.09 fb \\ 
 %    $Z,\gamma \to \tau\tau$ & 4.6 fb & 7.3 fb & 5.6 fb & 2.1 fb & 0.6 fb & 0.01 \\ \hline
 %   Total & 207.8 fb & 166.6 fb & 98.4 fb & 74.4 fb & 49.6 fb & 37.8 fb  \\ 
 \hline
    \end{tabular}
\caption{
  Background cross sections for $j_{\tau} \mu$ final state
  after cuts at PL.
%  See text for $j_{\tau}$ tagging efficiency and light-jet mistag rate.
%simplicity, the average $j_{\tau}$ tagging efficiency is set at $\epsilon_{\tau}$ = 0.7,
%  and rate of light jet mistagged as $j_{\tau}$ is taken as
%  1/35 for 1-prong and 1/240 for 3-prong.
}
 \label{crossx_slep_tm_bkgs}
\end{table}

\begin{table}[t]\centering
    \begin{tabular}{cll} \hline
    Variables & \ Leptonic & Semi-leptonic  \\ \hline
    $p_{T}(e)$  & \ $>$ 13 GeV & \ NA. \\
    $p_{T}(\mu)$ & \ $>$ 10 GeV & \ $>$ 30 GeV \\
    $p_T(j_{\tau})$ & \ NA & \ $>$ 25  GeV \\
    $|\eta(e)|$ & \ $<$ 2.5 & \ NA \\
    $|\eta(\mu)|$ & \ $<$ 2.4 & \ $<$ 2.5 \\
    $|\eta(j_{\tau})|$ \ & NA  & \ $<$ 2.3 \\
    $\slashed{E}_T$ & \ $>$ 20 GeV & \ $>$ 20 GeV \\
    $M_T(\mu, \slashed{E}_T)$ & \ $<$ 50 GeV & \ $<$ 50 GeV \\
    $M_{T}(e,\mu,\slashed{E}_T)$ & \ $(0.4\, m_H , 0.95\, m_H)$ & \ NA \\
     $M_{T}(j_{\tau},\mu,E_T)$& \ NA & \ $(0.4\, m_H , 0.95\, m_H)$ \\ 
    $\Delta R (e,\mu)$ & \ (3,4.5) & \ NA \\ 
    $\Delta R (\tau_j,\mu)$ & \ NA & \ $>$ 2.4 \\
    $N_j$ &  \ NA & \ 0 \\ 
    \hline
    \end{tabular}
\caption{Cuts for $H,A \to \tau\tau$.}
 \label{tab:cutstt}
%\label{tab:my_label}
\end{table}

\section{\boldmath
 Collider prospects for \large{$H,A \to \tau\tau$}
}

The $H,A \to \tau\tau$ process is more challenging 
than $H,A \to \tau\mu$ to probe at LHC because 
mass reconstruction is poorer.
The tau pairs from $gg \to H,A \to \tau\tau$ are back to back.
It is difficult to determine the invariant mass of tau pairs
($M_{\tau\tau}$).
Thus we rely on the transverse mass with visible particles and
missing transverse energy from tau decays.
Furthermore, for $\rho_{\tau\tau} =\rho_{\tau\mu} = \lambda_{\tau}$, 
the $H,A \to \tau\tau$ final state has half the event rate
compared to $H,A \to \tau^+\mu^- +\tau^-\mu^+$.
Although there is no flavor violation,
we consider the same final state as
$\tau\tau \to e\mu + \slashed{E}_T, \ j_\tau\mu +\slashed{E}_T$.

\vskip0.175cm
\noindent{\bf Contribution of $\tau\mu$ in $\tau\tau$.} 
With the same final state, obviously $H, A\to \tau\mu$ 
can contribute to $H, A \to \tau\tau$. 
However, we find that $M_{T}(\mu, \slashed{E}_T)$ 
is a powerful variable in separating 
the $\tau\tau$ signal from $\tau\mu$. 
This is because 
$M_T(\mu, \slashed{E}_T) >$ 100 GeV for $\tau\mu$, while 
$M_T(\mu, \slashed{E}_T) < 50 $ GeV for $\tau\tau$, 
as $\mu$ from the former comes from Higgs decay whereas 
for $\tau\tau$ it comes from $\tau$ decay.  

For $H, A \to \tau\tau$, both leptons 
(fully leptonic) and $j_{\tau}$ + lepton 
(semileptonic) come from $\tau$ decay, 
which give 4 and 3 neutrinos, respectively, 
in the final state. This makes the 
collinear approximation weaker for mass reconstruction. 
So we rely on cluster transverse mass of 
two visibly decaying $\tau$'s and $\slashed{E}_T$ 
 ($M_T(\tau_{\rm vis1},\tau_{\rm vis2},\slashed{E}_T)$), 
{which is given as~\cite{Barger:1987nn},}
\begin{align}
    M_T^2(\tau_{\rm vis1},\tau_{\rm vis2},\slashed{E}_T)
 = &\left(
        \sqrt{|\vec{p}_{T}(\tau_{\rm vis1},\tau_{\rm vis2})|^2
     +M^2(\tau_{\rm vis1},\tau_{\rm vis2})} + \slashed{E}_T
    \right)^2
 \nn \\ 
 &\ \ \ - \left(\vec{p}_T(\tau_{\rm vis1},\tau_{\rm vis2}) + \vec{\slashed{E}}_T\right)^2,
\end{align}
with $M(\tau_{\rm vis1},\,\tau_{\rm vis2})$ 
and $\vec{p}_{T}(\tau_{\rm vis1},\,\tau_{\rm vis2})$  
the invariant mass and net transverse momentum of 
the two visible $\tau$ decays, respectively.

\begin{figure*}[t]
    \centering
    \includegraphics[scale=0.15]{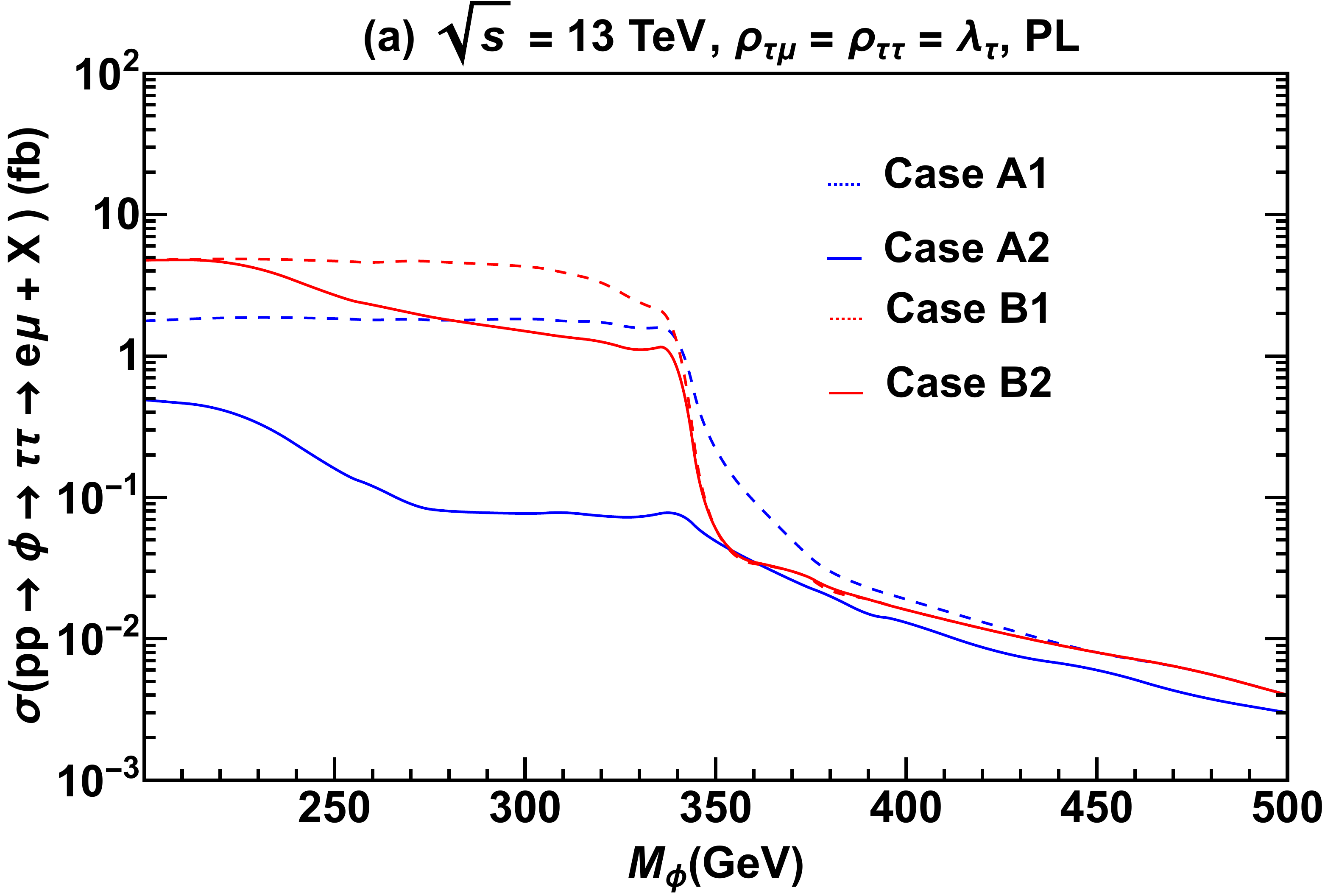}
    \includegraphics[scale=0.15]{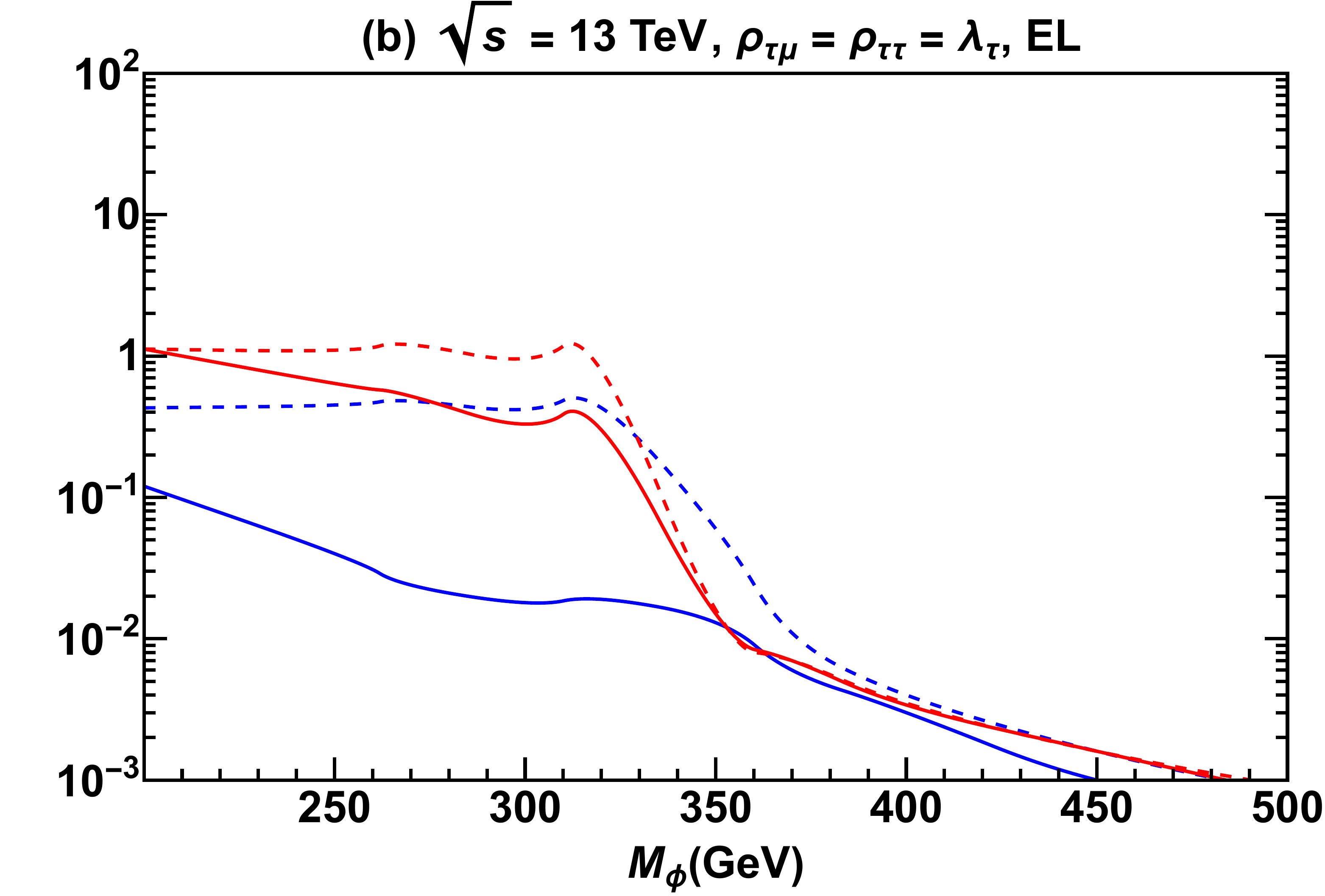}\\
    \includegraphics[scale=0.15]{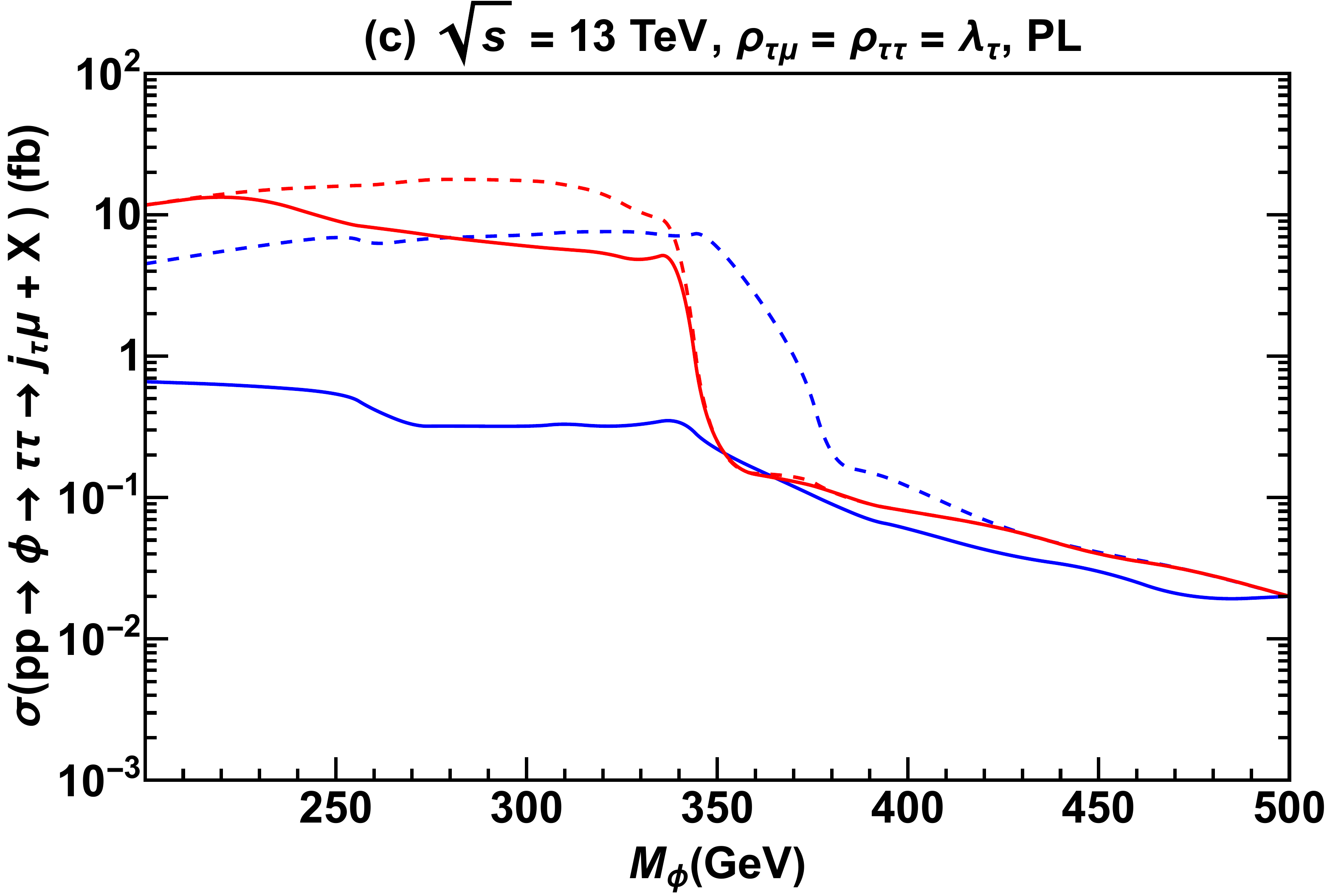}
    \includegraphics[scale=0.15]{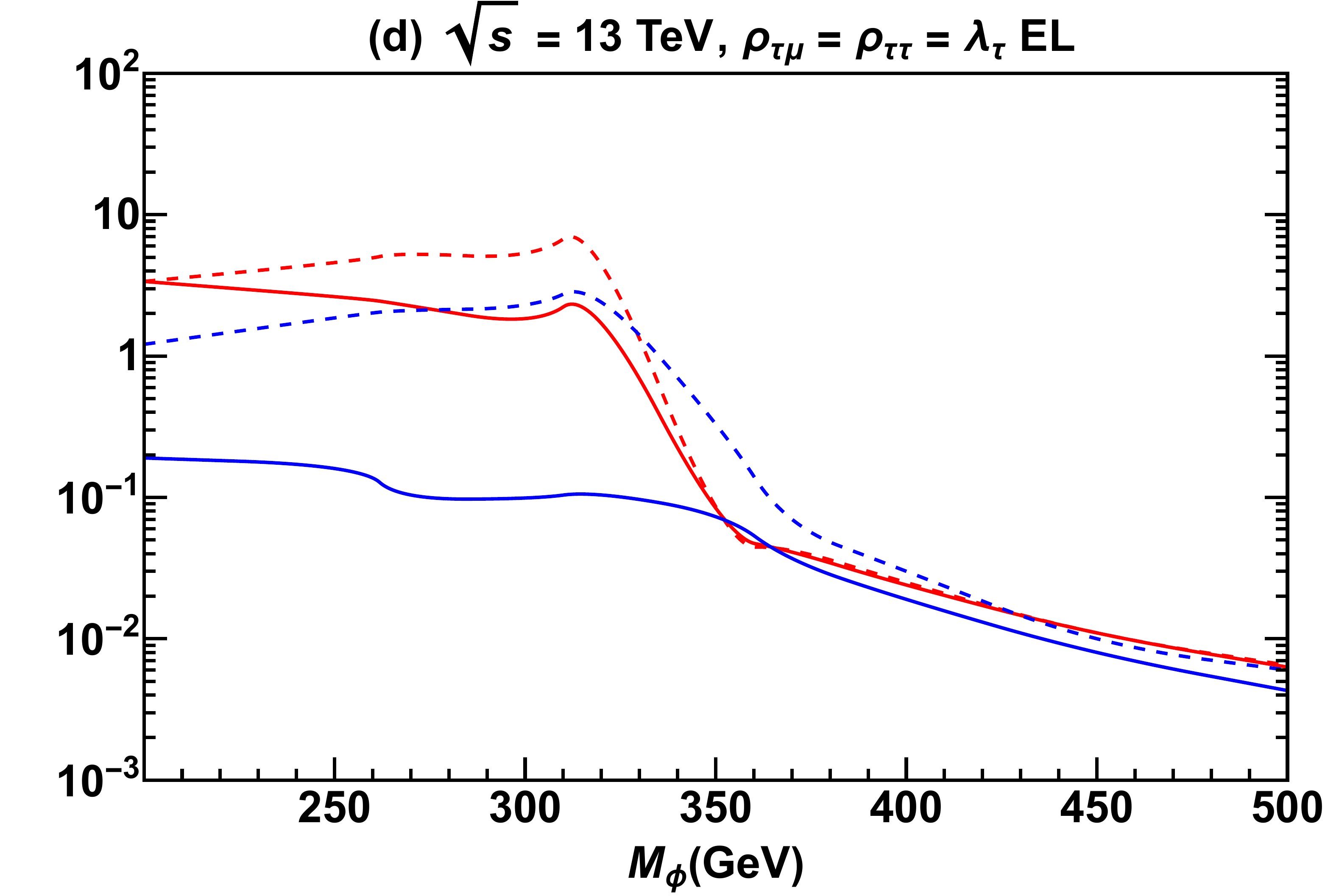}
\caption{
  Cross sections for $pp \to H,A \to \tau\tau$ for both leptonic (a, b)
  and semileptonic (c, d) final states after applying 
  all cuts at parton (a, c) and event (b, d) levels.}
    \label{fig:my_label}
\end{figure*}

Following ATLAS~\cite{ATLAS:HTATA}, our selection rules 
for $e\mu$ and $j_{\tau}\mu$ final states are given in Table~\ref{tab:cutstt}.
The cross section for the signal at both EL and PL 
for leptonic and semileptonic channels is presented 
in Figure~\ref{fig:my_label}. 
The background cross sections are given 
in Table~\ref{crossx_lep_tt_bkgs} after all cuts 
at both EL and PL for leptonic channel, 
and in Table~\ref{crossx_slep_tt_bkgs} 
only at PL for semileptonic channel.

\begin{table}[t]
    \centering
    \begin{tabular}{clllllll}
%    \multicolumn{5}{c}{$pp \to H,A \to \tau \tau$} \\ \hline
    \multicolumn{6}{c}{{\it Parton Level} ($pp \to H,A \to \tau \tau$)} \\ \hline
    Backgrounds/$m_H$ \ \ & 200 GeV &\ 250 GeV & \ 300 GeV & \ 350 GeV & \ 400 GeV & \ 450 GeV & \ 500 GeV \\ \hline \hline
     $Z,\gamma \to \tau\tau$ & \ 195.0 fb & \  38.6 fb & \ 23.1 fb & \ 15.3 fb & \ 10.5 fb & \ \ 7.6 fb & \ \ 5.6 fb \\
    $W^+W^-$ & \ \ 41.2 fb &  \ 52.5 fb & \ 53.3 fb & \ 48.3 fb & \ 40.7 fb & \ 32.9 fb & \ 26.1 fb  \\
    $t\bar{t}$ & \ \ 3.60 fb & \ \ 4.5 fb & \ \ 4.8 fb & \ \ 4.7 fb & \ \ 4.3 fb & \ \ 3.9 fb & \ \ 3.4 fb \\
    $tW^{\pm}$ &  \ \ 2.66 fb &\ \ 6.0 fb & \ \ 6.6 fb & \ \ 6.6 fb & \ \ 6.2 fb & \ \ 5.6 fb & \ \ 5.0 fb \\
    
     \hline
    Total  &  \ 243.5 fb&  101.6 fb &  \ 87.8 fb & \ 74.8 fb & \ 61.8 fb & \ 50 fb & \ 40.0 fb \\ \hline \hline
    \multicolumn{6}{c}{{\it Event Level} ($pp \to H,A \to \tau \tau$)} \\ \hline
    $Z,\gamma \to \tau\tau$ & 71.7 fb & \ 14.2 fb & \ \ 9.3 fb & \ \ 6.6 fb & \ \ 4.6 fb & \ \ 3.4 fb & \ 2.6 fb \\
    $W^+W^-$ & 11.8 fb & \ 15.0 fb & \ 15.9 fb & \ 14.9 fb & \ 13.2 fb &  \ 11.1 fb & \ 9.1 fb  \\
    $t\bar{t}$ & \ 1.9 fb &\ \ 2.4 fb & \ \ 2.8 fb & \ \ 3.1 fb & \ \ 3.1 fb & \ \ 3.0 fb & \ 2.7 fb \\
    $tW^{\pm}$ & \ 0.9 fb  &\ \ 2.1 fb & \ \ 2.6 fb & \ \ 2.7 fb & \ \ 2.6 fb & \ \ 2.4 fb & \ 2.1 fb \\ \hline
    Total & 86.3 fb & \ 33.7 fb & \ 30.5 fb & \ 27.3 fb & \ 23.4 fb & \ 19.9 fb & 16.5 fb  \\ \hline
    \end{tabular}
 \caption{
  Background cross sections for $e \mu$ final state 
  after cuts at PL and EL in $\tau\tau$ channel.
}
 \label{crossx_lep_tt_bkgs}
\end{table}

\begin{table}[t]
    \centering
    \begin{tabular}{clllllll}
%    \multicolumn{5}{c}{$pp \to H,A \to \tau \tau$} \\ \hline
    \multicolumn{6}{c}{{\it Parton Level} ($pp \to H,A \to \tau \tau$)} \\ \hline
    Backgrounds/$m_H$ \ \ & \ \ 200 GeV &\ \ 250 GeV & \ \ 300 GeV & \ \ 350 GeV & \ \ 400 GeV & \ \ 450 GeV & \ \ 500 GeV \\ \hline \hline
    $W^{\pm}j$ & 2790.9 fb & 3463.7 fb &  \; 3515.4 fb & \ 3215.5 fb & \ 2794.3 fb & \ \ 2365 fb & \ 1976.1 fb  \\
    $Z,\gamma \to \tau\tau$ &\ \ \ 97.9 fb &\ \ 118.7 fb & \ \ \  101.1 fb & \ \ \ \ 81.2 fb & \ \ \ \  63.6 fb & \ \ \ 49.6 fb & \ \ \ 38.8 fb \\
    $t\bar{t}$ &\ \  \ 44.8 fb  &\ \ \hspace{0.5mm} 78.3 fb & \ \ \ \hspace{0.2mm}  99.9 fb & \ \ \ 113.2 fb & \ \ \ 112.2 fb & \ \ 106.6 fb & \ \ 96.3 fb \\
    $tW^{\pm}$ &\ \ \ \ 8.1  fb &\ \ \hspace{0.5mm} 12.9 fb & \ \ \ \ \ 16.6 fb & \ \ \ \ 18.6 fb & \ \ \ \ 19.2 fb & \ \ \ \ 18.8 fb & \ \ \ 17.6 fb \\ 
    $W^+W^-$ & \ \ \ \ 6.6  fb & \ \ \ \ \ 8.4 fb & \ \ \ \ \  \hspace{0.5mm} 8.7 fb & \ \ \ \ \   8.3 fb & \ \ \ \ \ 7.6 fb & \ \ \ \ \ 6.8 fb & \ \ \ \ 6.1 fb \\ \hline
    Total & 2948.3 fb & \ 3681.1 fb &  \ 3740.2 fb & 3435.3 fb & 2997.0 fb & 2547.5 fb & 2133.7 fb \\ \hline \hline
    %\multicolumn{5}{c}{{\it Event Level} ($pp \to H,A \to \tau \tau$) \\ \hline
   %  $W^{\pm}j$ & 681.7 fb & 780.9 fb & 760.9 fb & 680.1 fb & 585.4 fb & 454.6 fb  \\
  %  $W^+W^-$ & 0.58 fb & 0.82 fb & 0.98 fb & 1.01 fb & 0.92 fb & 0.87 fb  \\
  %  $Z,\gamma \to \tau\tau$ & 14.2 fb & 9.3 fb & 6.6 fb & 4.6 fb & 3.4 fb & 2.6 fb \\
  %  $t\bar{t}$ & 4.04 fb  & 6 fb & 7.4 fb & 7.9 fb & 7.8 fb & 7.3 fb \\
  %  $tW^{\pm}$ & 0.04 fb & 0.11 fb & 0.12 fb & 0.12 fb & 0.1 fb & 0.08 fb \\ \hline
  %  Total & 33.7 fb & 30.5 fb & 27.3 fb & 23.4 fb & 19.9 fb & 16.5 fb  \\ 
  \hline
    \end{tabular}
\caption{
  Background cross sections for $j_{\tau} \mu$ final state after cuts
  at PL in $\tau\tau$ channel.}
 \label{crossx_slep_tt_bkgs}
\end{table}

%-----------------------------------------
%  Section # 6
%------------------------------------------
\section{Statistical Significance of the Signal}

We now estimate the discovery potential 
of all channels we have discussed. We have 
kept $\rho_{\tau\mu} = \rho_{\tau\tau} = \lambda_{\tau}$ 
in Figs.~\ref{fig:crossxtm} and \ref{fig:my_label} 
for simplicity. In this section, however, 
we consider the constraints on $\rho_{\tau\mu}$ 
and $\rho_{\tau\tau}$ as discussed in 
Figs.~\ref{fig:lim_rho_lm} and~\ref{fig:lim_rho_ll}. 
We scale our signal cross section for $\tau\mu$ channel 
using the most strict limit for each case. 
For the $\tau\tau$ channel, especially with 
$h\to \tau\tau$ constraint coming into play, we
use $\rho_{\tau\tau} < 0$ limits to enhance 
the signal estimates for Cases A1 and A2, 
as $\lambda_{H\tau\tau} \simeq \lambda_{\tau} \rm c_{\gamma}
 - \rho_{\tau\tau} \rm s_{\gamma}$.\footnote{
 We set $\rm s_{\gamma}$ as positive.
} 
For Cases B1 and B2, $A \to \tau \tau $ limits 
are more stringent for $m_A < 2 m_t$, 
and beyond which we choose {the magenta dashed
 (CMS +1$\sigma$)} of Fig.~\ref{fig:lim_rho_ll}
to stay within experimental constraints.

\begin{figure*}[t]
    \centering
    \includegraphics[scale=0.14]{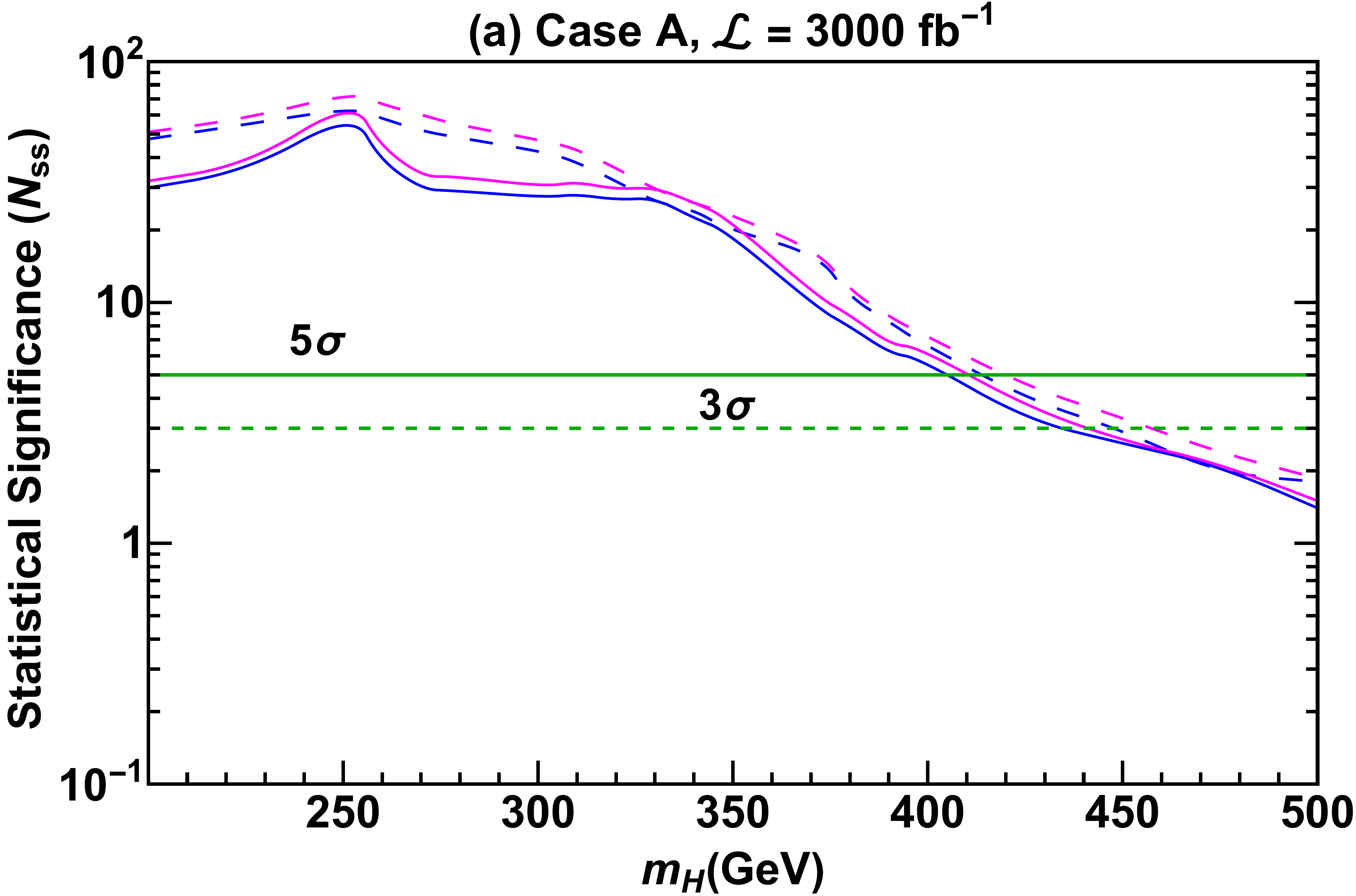}
    \includegraphics[scale=0.14]{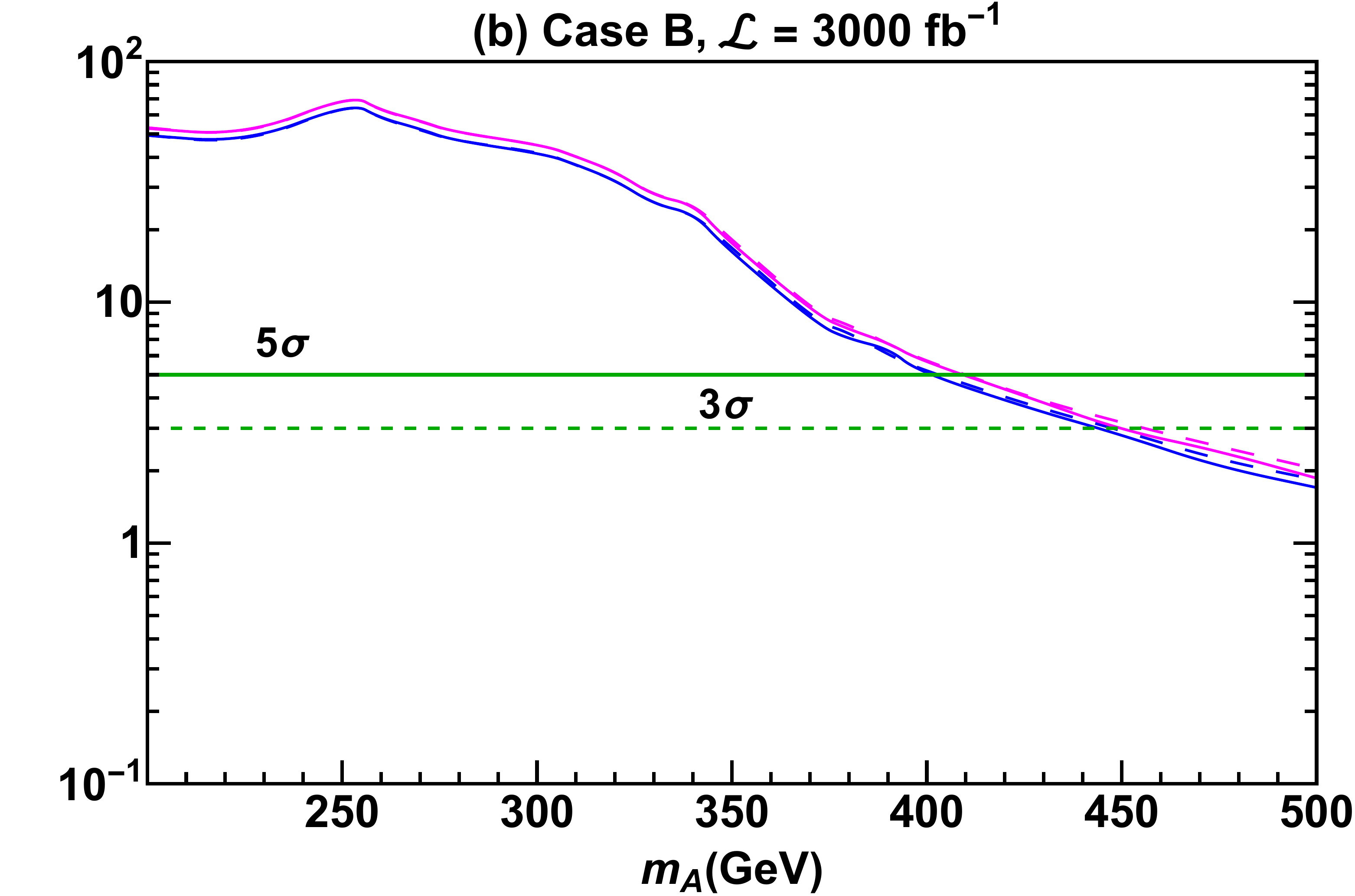}\\
    \includegraphics[scale=0.14]{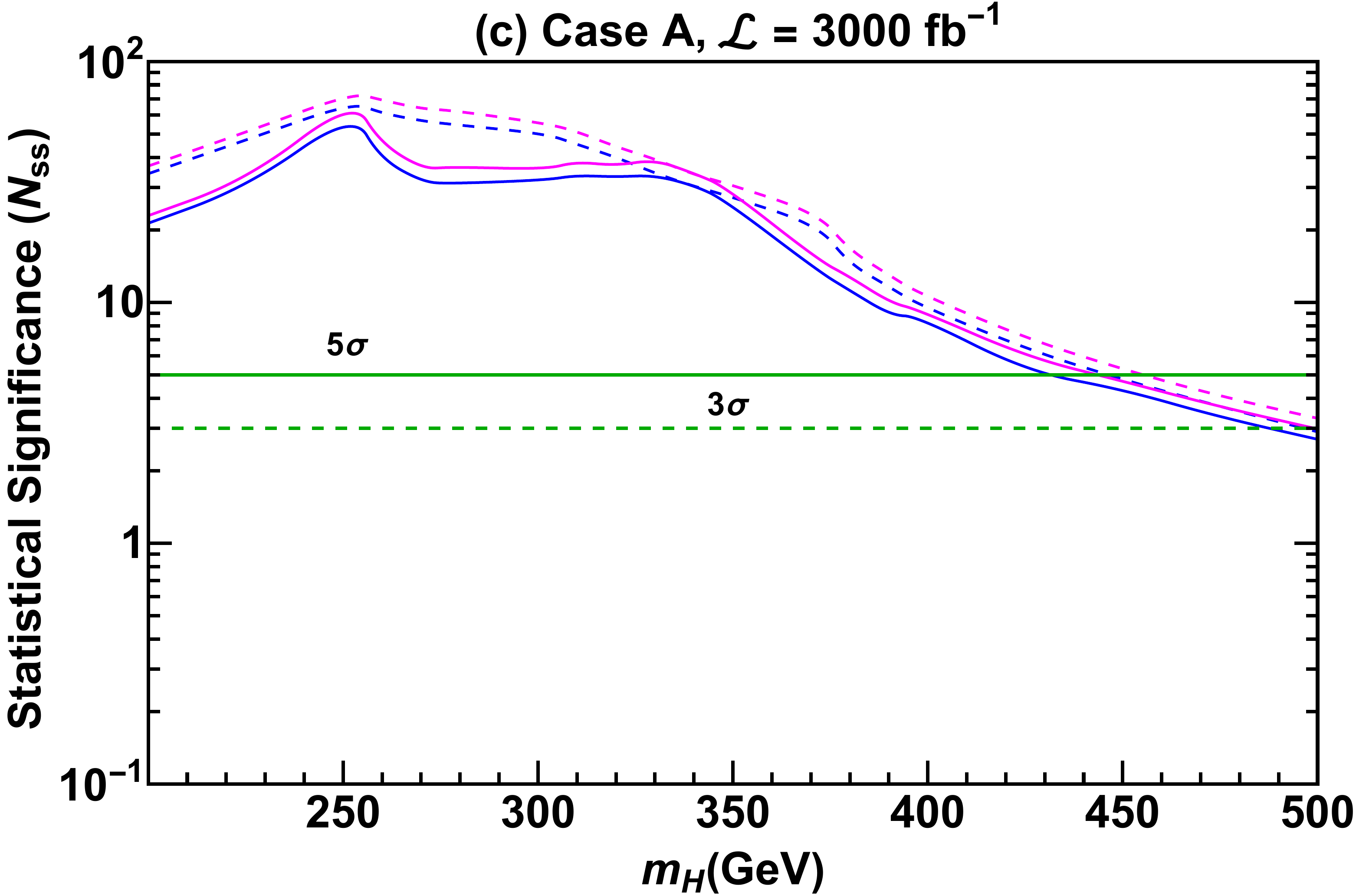}
    \includegraphics[scale=0.14]{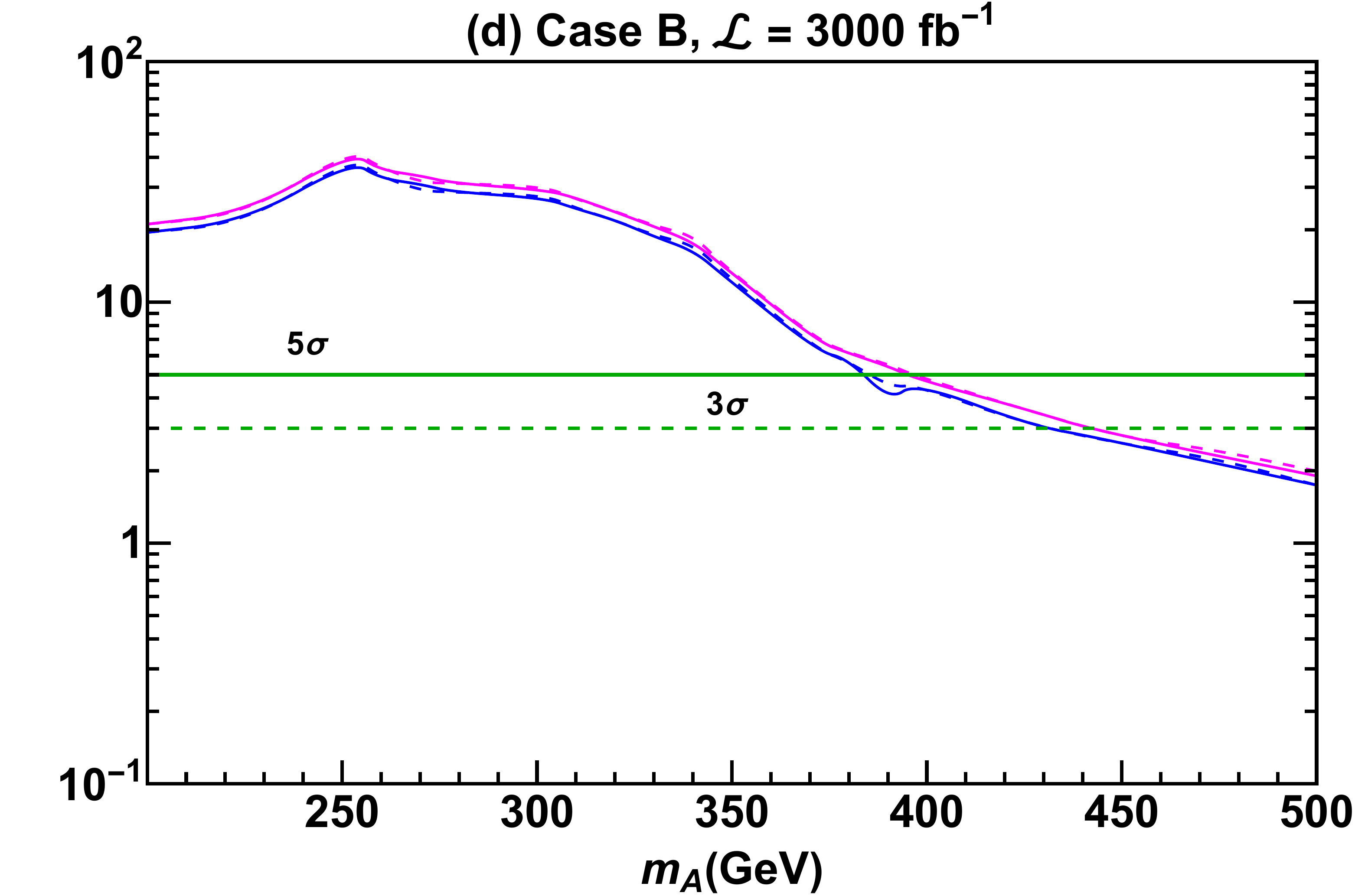}
\caption{
  Statistical significance $N_{SS}$ for $p p \to H,A \to \tau \mu$ at PL, where 
  (a), (b) are for the fully leptonic channel,  and (c), (d) for the semileptonic channel. 
  Both $\sqrt{s}$ = 13 (blue) and 14 TeV (magenta) are given, 
  where solid (dashed) lines are for $\cos\gamma$ = 0.01 (0.1). 
  The left panels (a, c) are for $m_H < m_A = m_{H^+}$, 
  and (b, d) for $m_A < m_H = m_{H^+}$. We have used $\rho_{\tau\mu} = \rho_{\mu\tau} =$ limits, derived in Section III, i.e. Fig. 4. 
}
\label{fig:signi_PL_lm}
\end{figure*}

% {\tcr { To estimate the discovery potential, we estimate the lower limit on 

\begin{figure*}[t]
    \centering
    \includegraphics[scale=0.135]{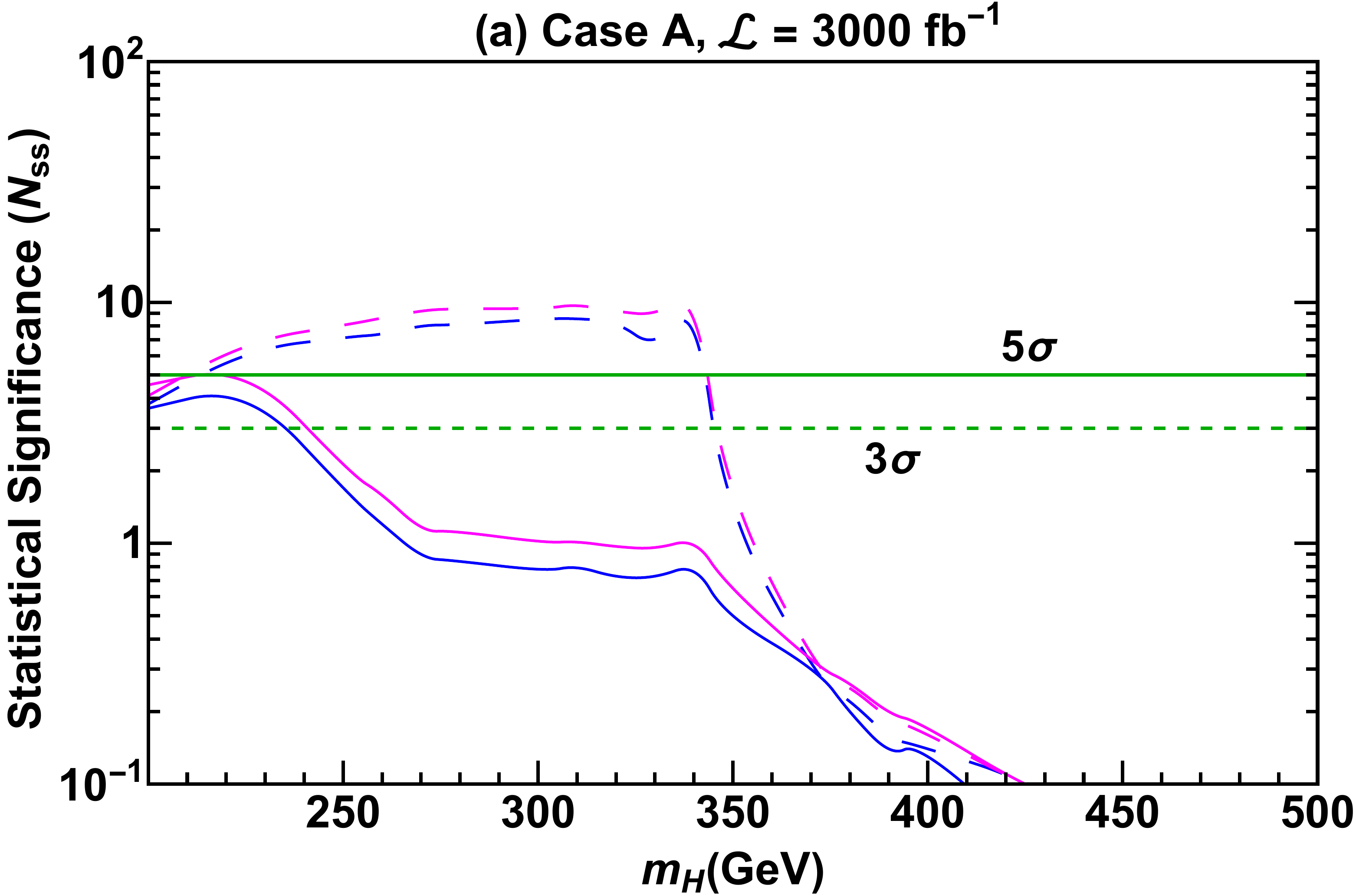}
    \includegraphics[scale=0.135]{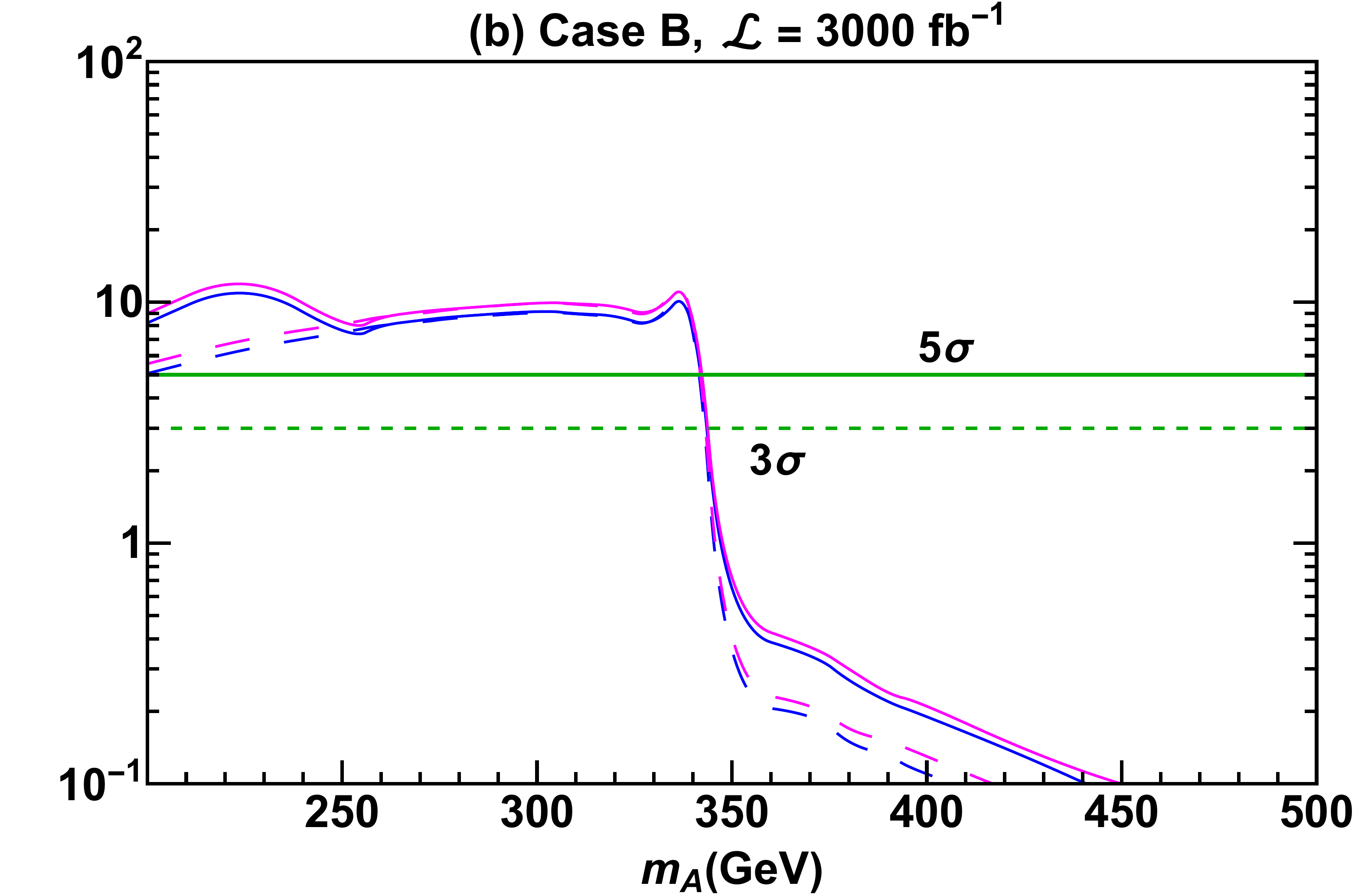}\\
    \includegraphics[scale=0.135]{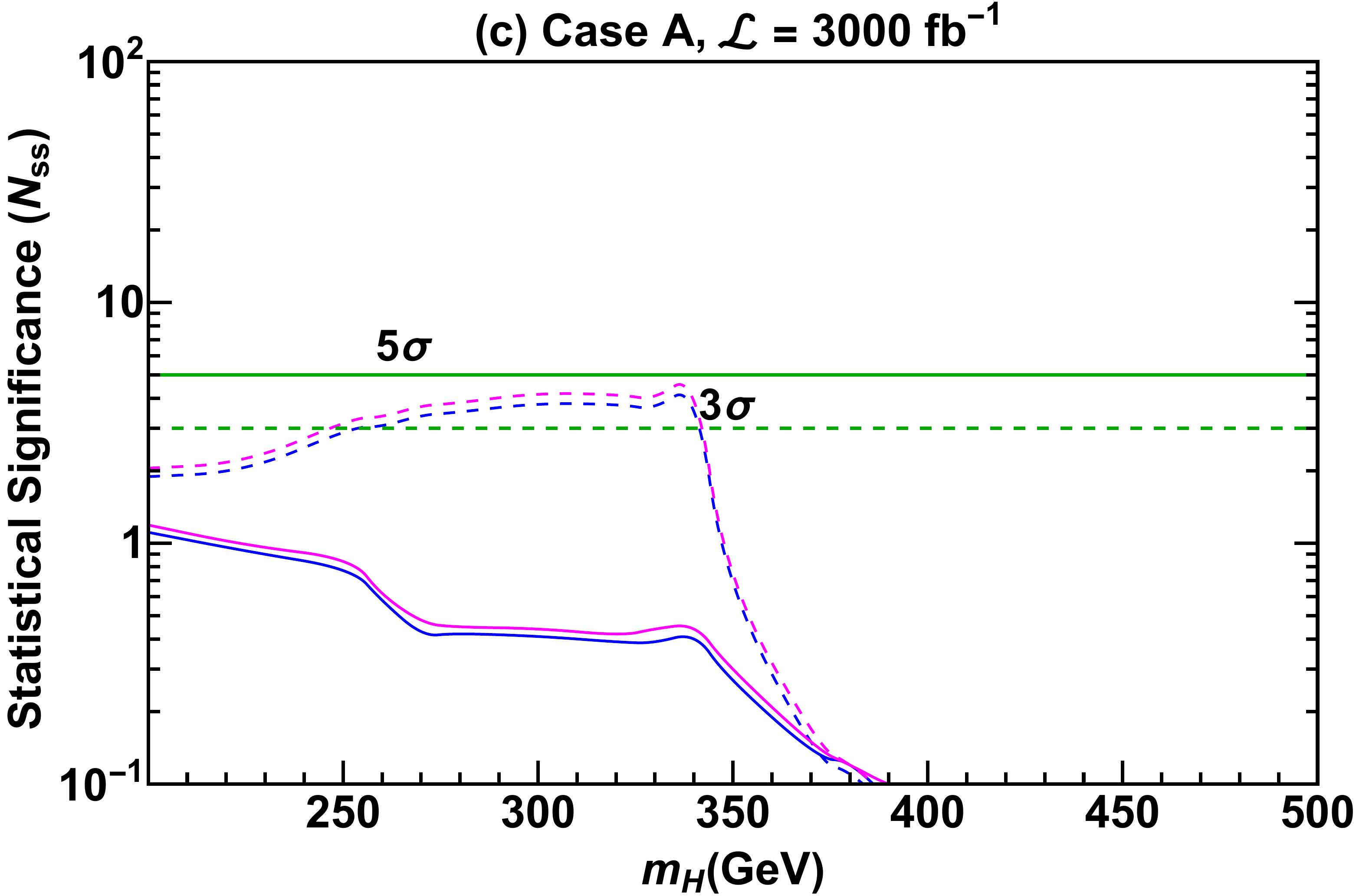}
    \includegraphics[scale=0.135]{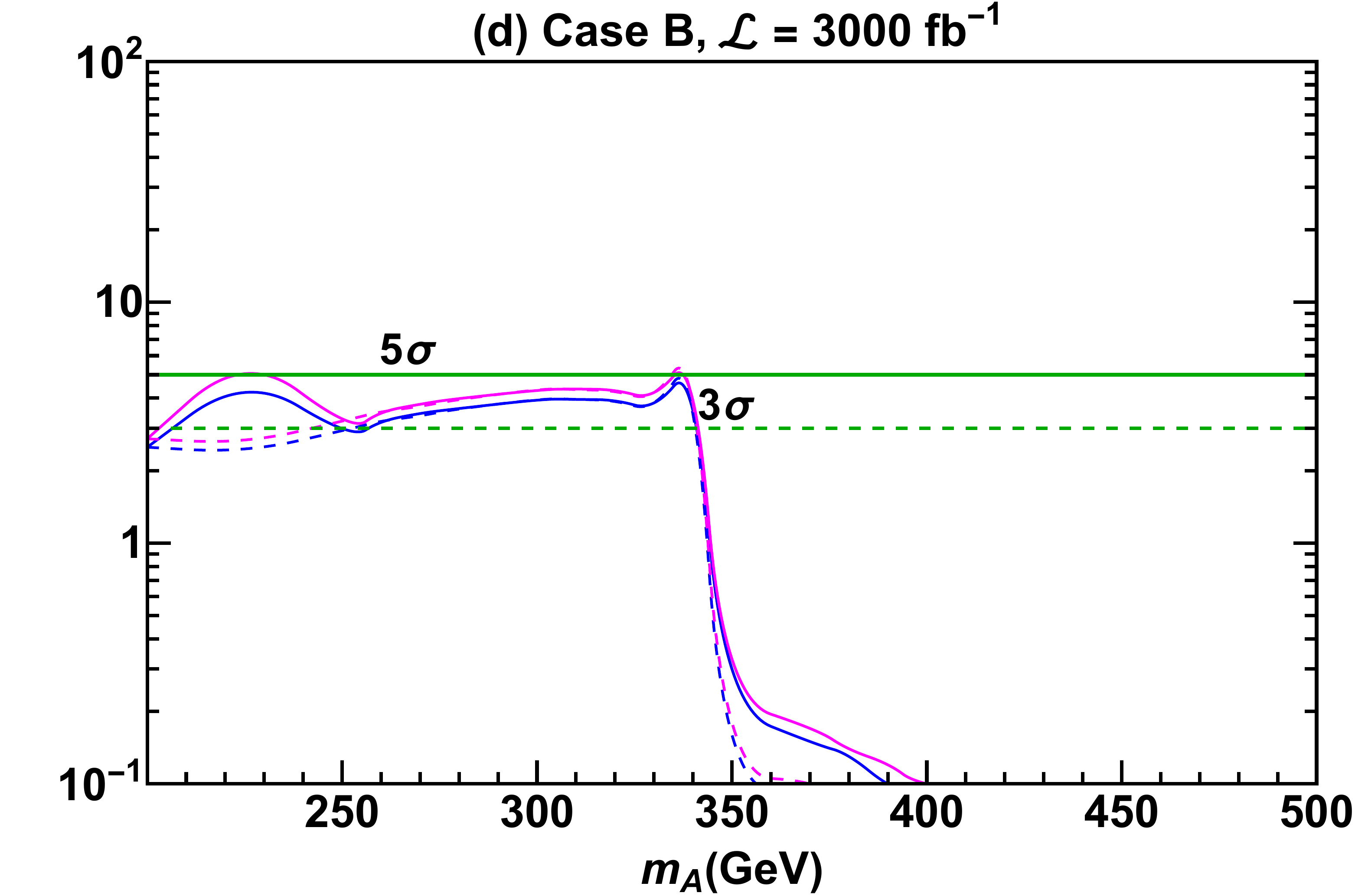}
\vskip-0.2cm
\caption{
  Same as Fig.~\ref{fig:signi_PL_lm} for $p p \to H,A \to \tau \tau$ at PL. 
}
\label{fig:signi_PL_ll} 
\end{figure*}

To estimate significance, we assume Gaussian distribution,
and denote $N_S$ as the number of signal events, 
and $N_B$ for background events
 (combining all background processes).
The statistical significance $N_{SS}$ is evaluated with~\cite{Cowan:2010js} 
\begin{equation}
N_{SS} = \sqrt{2 (N_B + N_S) \rm \ln(1 + \frac{N_S}{N_B}) - 2 N_S} \, .
\label{eq:signif}
\end{equation}
For a large number of background events ($N_B \gg N_S$),
it simplifies to become the well known discovery significance
\begin{equation}
  N_{SS}
  = \frac{N_S}{\sqrt{N_B}} \, .
\end{equation}
We estimate the statistical significance for 
each signal point using Eq.~(\ref{eq:signif}) 
at the parton level for $H,A\to \tau\mu$ 
and $H,A \to \tau \tau$, for both fully 
leptonic and semileptonic channels. 
We present our estimates for $N_{SS}$ 
at PL for $H,A \to \tau\mu$ and $H,A \to \tau \tau$, 
respectively, in Figs.~\ref{fig:signi_PL_lm} 
and \ref{fig:signi_PL_ll} for $\sqrt{s} =$ 13 and 14 TeV 
at $\mathcal{L} = 3\, \rm ab^{-1}$. 
In both figures, we give $N_{SS}$ for the purely leptonic channel in (a) and (b), 
and for semileptonic channels in (c) and (d).

\begin{figure*}[h]
    \centering
    \includegraphics[scale=0.135]{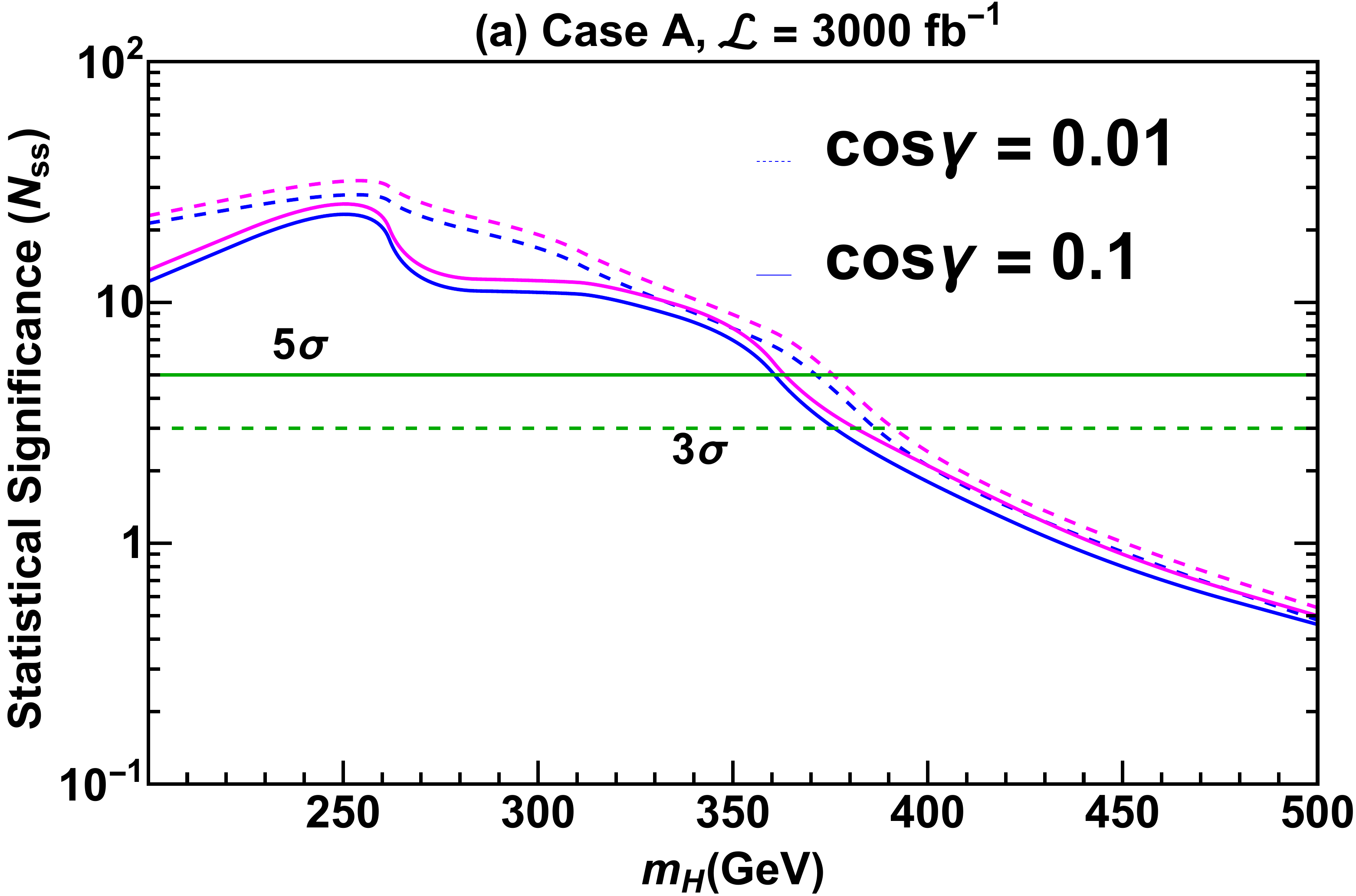}
    \includegraphics[scale=0.135]{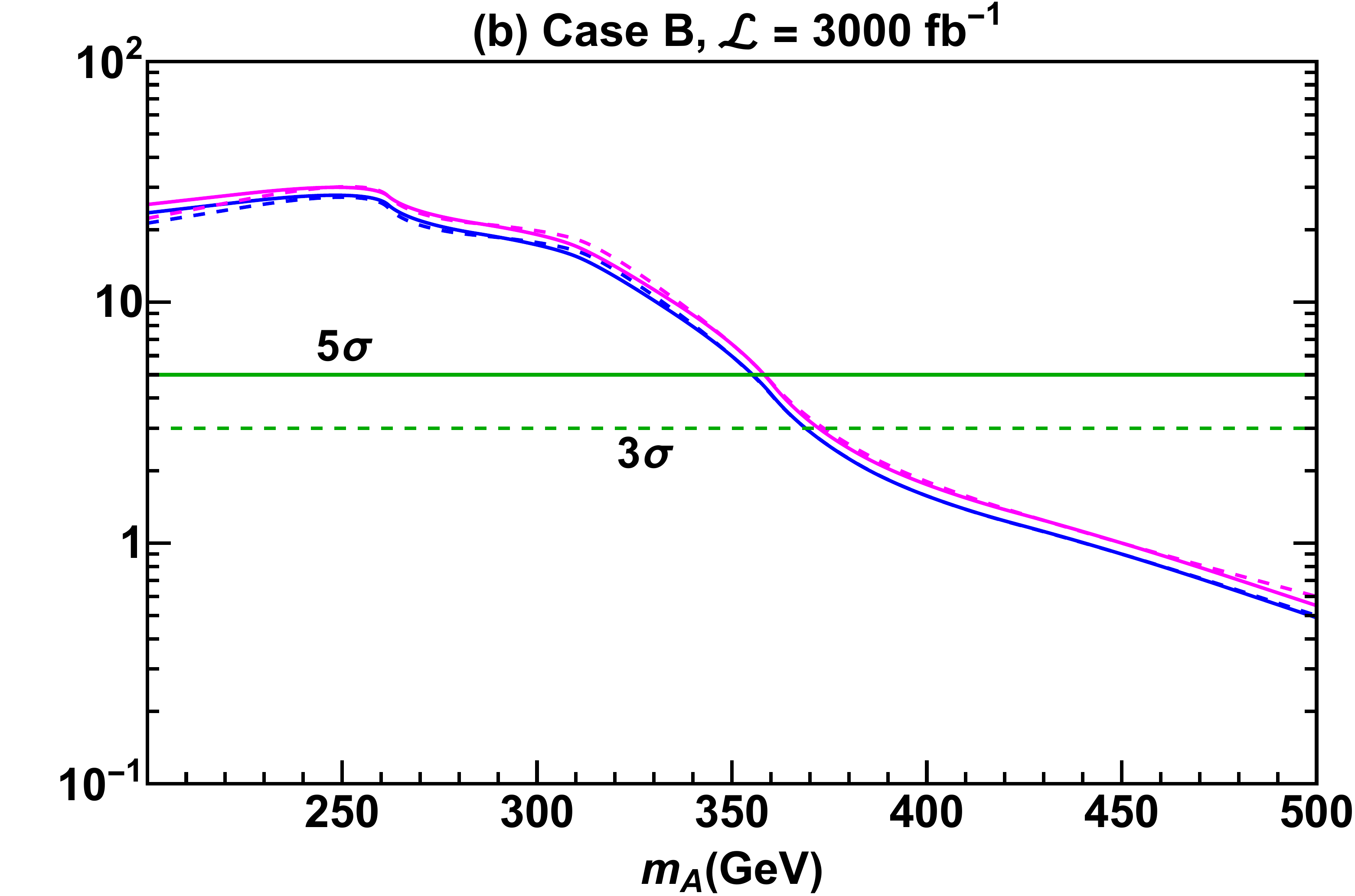} \\
    \includegraphics[scale=0.135]{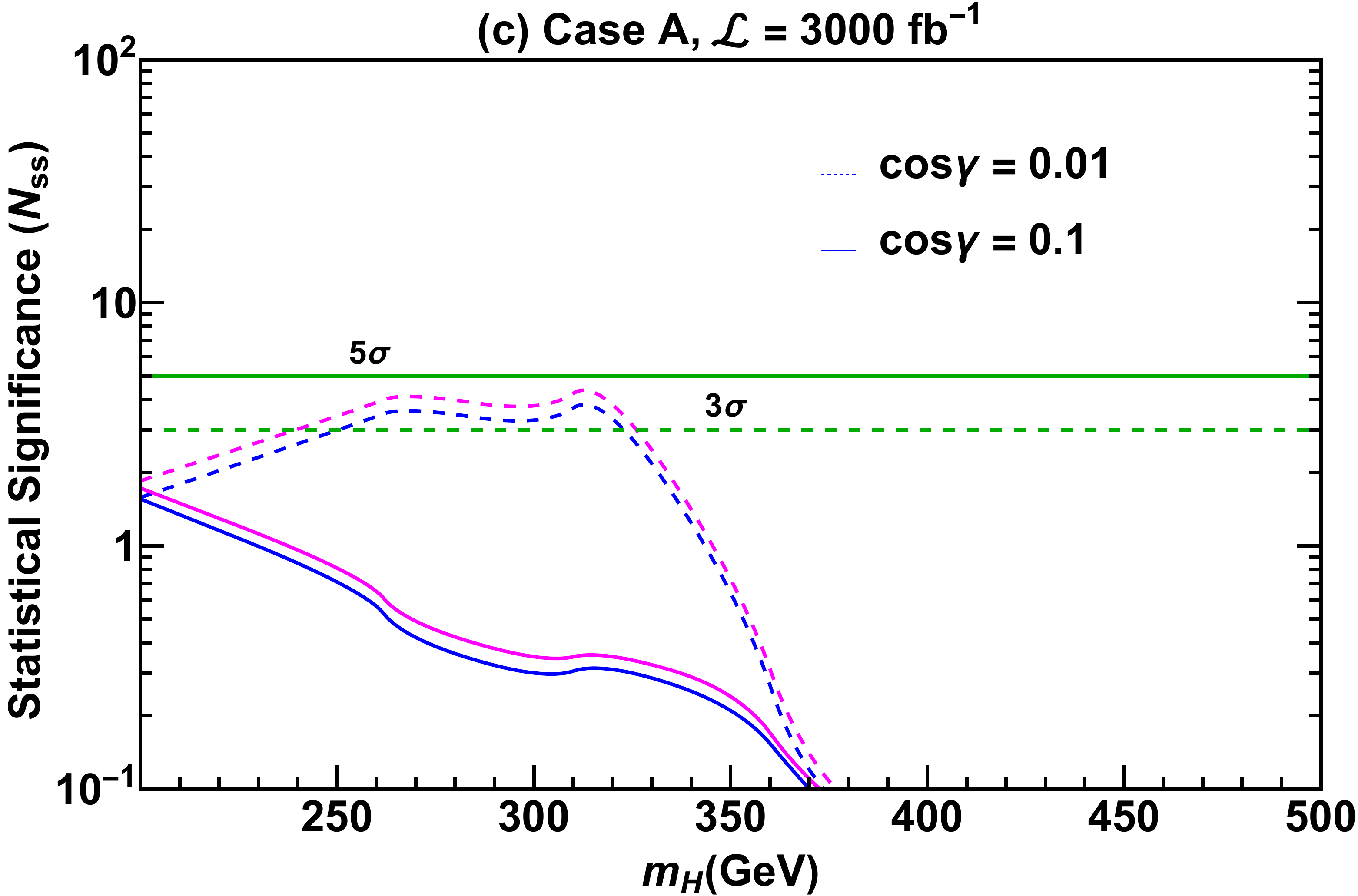}
    \includegraphics[scale=0.135]{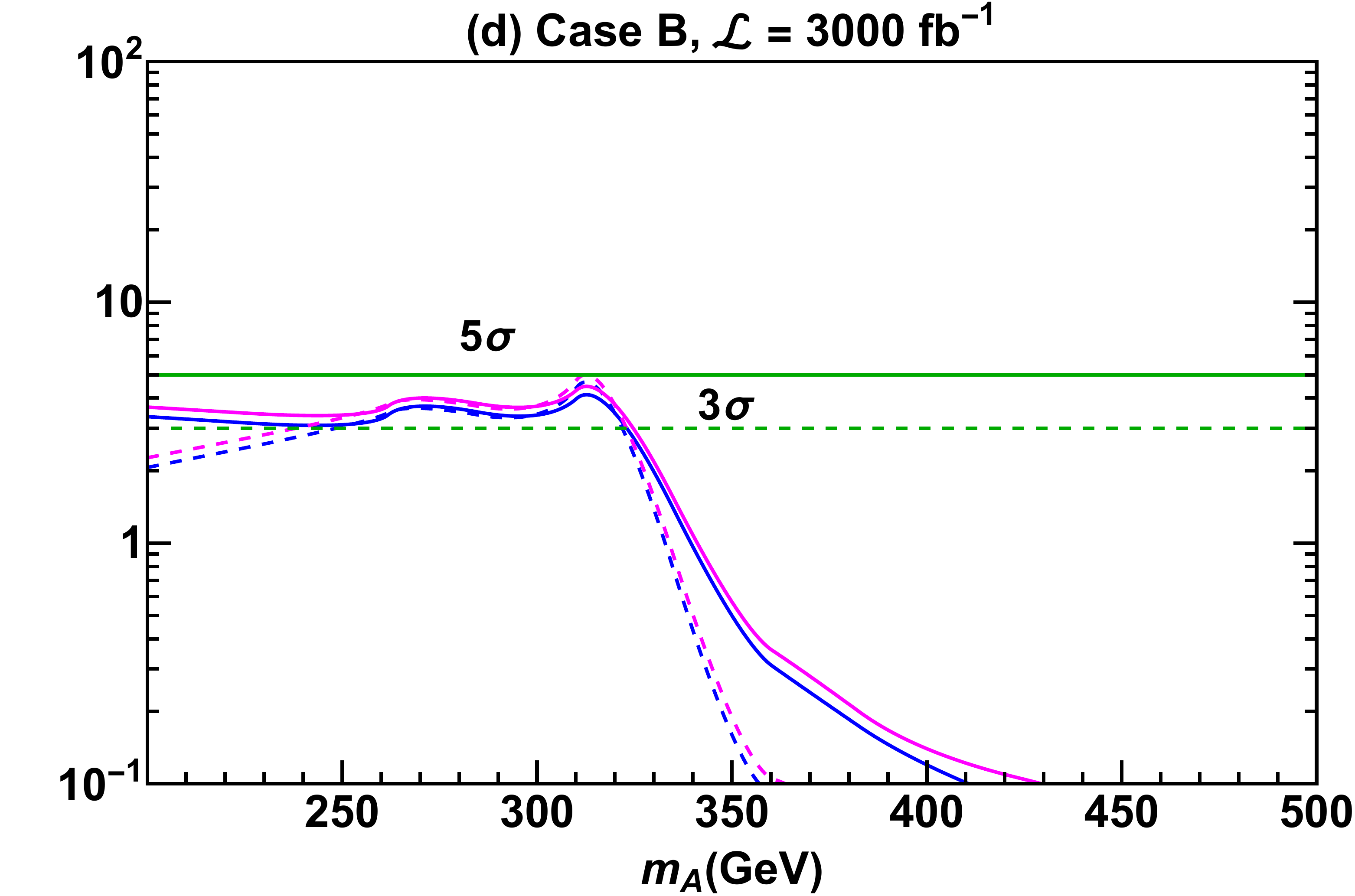}
\vskip-0.2cm
\caption{
  Statistical significance $N_{SS}$ at the event level
  for $p p \to H,A \to \tau \mu \to e \mu +\slashed{E}_T +X$ (top row)
  and $p p \to H,A \to \tau \tau \to e mu +\slashed{E}_T +X$ (bottom row).   
  Both $\sqrt{s}$ = 13 (blue) and 14 TeV (magenta) are presented, 
  where solid (dashed) lines are for $\cos\gamma$ = 0.01 (0.1). 
  The left panels (a, c) are for $m_H < m_A = m_{H^+}$, 
  and (b, d) for $m_A < m_H = m_{H^+}$.
  We have chosen $\rho_{\tau\mu} = \rho_{\mu\tau} =$ limits,
  derived in Section III, i.e. Fig. 4. 
}
\label{fig:my_sig_EL}
\end{figure*}

We only present $N_{SS}$ for purely leptonic channel  at EL for simplicity, 
%{\color{red} This is mainly due to uncertainty in connecting the PL analysis with EL analysis for semi-leptonic channel. At PL we randomly mis-tag a light jet for all the semi-leptonic backgrounds except $Z,\gamma \to \tau\tau$ and veto event with more than one mis-tagged light jets. Since at PL there are no additional jets coming from Hadronization and cascade decays it becomes difficult to connect the PL with EL}. 
%
which is given in Fig.~\ref{fig:my_sig_EL} 
for $\tau\mu$ in (a) and (b), 
and for $\tau\tau$ in (c) and (d).
%{\bf Event level:}

\section{Discussion and Conclusion}

Extra $\tau$ couplings $\rho_{\tau\tau}$ and $\rho_{\tau\mu}$ 
can act as good probes for exotic $H$, $A$ scalars
below the $t\bar{t}$ threshold at HL-LHC. 
We have illustrated the prospects of discovering 
either $H$ or $A$ in $\tau\mu$ and $\tau\tau$ final states. 
We studied the constraint on relevant couplings 
from various searches and estimated 
the statistical significance at the HL-LHC with
$3\, \rm ab^{-1}$ for $\sqrt{s} = 13$, and 14 TeV. 

From our study, we offer the following comments.
\begin{itemize}
\item 
CMS $H \to \tau \mu$ with $35.9 \, \rm fb^{-1}$ data 
puts stronger bound on $\rho_{\tau\mu}$ than 
the latest Belle limit on $\tau\to \mu \gamma$, which is 
limited to the mass range considered in this study. 
Intuitively, if $H$ or $A$ approach 
$\mathcal{O}(\rm TeV)$, we expect Belle to do better. 
Limits from $h \to \tau\mu$ depends on $c_{\gamma}$, 
strengthening as $c_{\gamma}$ increases from 0.01 to 0.1.
\item 
Constraints on $pp \to H \to \tau\tau$ 
from ATLAS and CMS follow an interesting trend: 
ATLAS is better for $m_H < 330$ GeV, 
but after which CMS and ATLAS are comparable. 
This again tells the amazing sensitivity of CMS, 
as the data used is only a quarter that of ATLAS. 
Constraints from $pp \to h \to \tau\tau$ 
become important for $c_{\gamma} = 0.1$. 
\item 
If $A$ is lighter, like in Cases B1 and B2, 
we find that the limits on $\rho_{\tau\mu}$ are 
the most stringent below $2m_t$, but 
becomes even weaker than Cases A1 and A2 
($H$ lighter) beyond $2m_t$. 
This is mainly due to $\Gamma(A\to t\bar{t})
 > \Gamma(H \to t \bar{t})$ with same mass. 
For all cases, the limits from CMS is better than 
$\rho_{\tau\mu} = \lambda_{\tau}$ 
at and below $2m_t$ threshold.
\item 
From our PL study of $pp \to H,A \to \tau \mu$, 
we offer some insight:
 \begin{itemize}
 \item
 Once we apply the $\rho_{\tau\mu}$ constraints,
 Cases B1 and B2 become almost identical.
 \item
 For Cases A1 and A2, lower value of $c_{\gamma}$
 does provide better significance, but above $2m_t$
 they become pretty close to each other. 
  \item
  We see a slight bump for Case A2 around $2m_t$,
  which is again a reflection of the limits that 
  we see in Fig.~\ref{fig:lim_rho_lm} as well as the rise in $pp \to H$ cross section at $t\bar{t}$ threshold.
  \item
 We find that below 400 GeV, $H$ or $A$ can
 be discovered by HL-LHC with just a single channel.
 Above 400 GeV, significance can be improved by combining leptonic and semileptonic channels. 
 \end{itemize}
\item 
The $pp \to H,A \to \tau \tau$ channel is much more challenging,
owing to poor mass reconstruction for $pp \to H,A \to \tau\tau$ and lower
branching fraction than $\tau\mu$ with $\rho_{\tau\tau} \simeq \rho_{\tau\mu}$.
We draw the following remarks at the parton level:
 \begin{itemize}
 \item Just like Cases B1 and B2 for $\tau\mu$,
 the statistical significance overlaps. We see an
 upward bump at the $t\bar t$ threshold, mainly due to the rise in production cross section.
 Beyond $2m_t$, the sharp rise of $t\bar{t}$
 kills the $\tau\tau$ channel.
 \item For Cases A1 and A2, $c_{\gamma}$ dependence 
 is clearly visible,
 and we prefer lower $c_{\gamma}$ for better significance.
 \item The semileptonic channel has a lower significance 
 due to high QCD background compared to 
 the much cleaner leptonic channel.
 \item HL-LHC can still discover this channel up to $t\bar{t}$ threshold,
 beyond that, we need much smarter classification techniques.
 \item Unlike $\tau\mu$ where we see 
 less steep a fall in statistical significance,
 for $\tau\tau$ there is a sharp drop in significance 
 after $2m_t$.
 This is mainly due to the limits on $\rho_{\tau\tau}$, 
 which stay almost the same for nearly 
 the entire mass range for all four cases. 
 \end{itemize}
\item At the event level, the statistical significance 
follows a similar trend as PL for the $e\mu$ channel 
as discussed. However, we see a drop in significance due to detector resolution 
and hadronization effects. 
But we can still discover the $\tau\mu$ channel 
below $2m_t$, and hopefully, by combining with 
the semileptonic channel, the discovery range can be 
further extended. However, $\tau\tau$ remains challenging. 
\end{itemize}
%
%What to discuss here?
% tc suppression, and implications on other channel
% tau-mu and tau tau mediated EWBG => Ramsey
% Exitence of combined rho_tc and rho_{\tau\mu} exciting new channels
% Beyond 2m_t is challenging and ML techniques can provide a way out
% More CMS data for H-> Tau mu and HL-LHC discovery might be next door => exciting time ahead.

As discussed in Sec.\;III, we have set $\rho_{tc} = 0$. 
A nonzero $\rho_{tc}$ can further dilute both signals. 
In Ref.~\cite{Hou:2019grj}, we showed that 
$\rho_{tc} \sim 0.5$ can dominate the 
branching fraction even beyond the $t\bar{t}$ threshold. 
However, we will need a very detailed study 
on overall impact of $\rho_{tc} \neq 0$, 
because it would relax the constraints from 
$H \to \tau \mu$ on $\rho_{\tau\mu}$ 
and from $H,A\to \tau\tau$ on $\rho_{\tau\tau}$. 
As discussed in Ref.\;\cite{Hou:2021xiq}, 
for the mass range considered here, 
$\rho_{tc} \sim 0.5$ might be too high and 
a more reasonable value would be $\rho_{tc} \sim 0.1$. 
But in that study, all $\rho_{ij}$s 
except $\rho_{tc}$ are set to zero,
and $c_{\gamma} = 0$ was taken.

Another big motivation to study 
$\rho_{\tau\tau}$ and $\rho_{\tau\mu}$ is 
driven by electroweak baryogenesis (EWBG). 
As discussed in Ref.~\cite{Guo:2016ixx,Ge:2020mcl}, 
complex extra $\tau$ couplings can also drive EWBG 
and explain matter-antimatter asymmetry of the Universe,
although it has been questioned~\cite{Cline:2021dkf} 
whether light fermions can actually achieve this. 
The $h, H, A \to \tau\tau$ processes are also 
considered as good channels to study CP violation~\cite{CMS:2021sdq,Antusch:2020ngh}, 
a necessary condition for EWBG.  
On the other side, there is EWBG driven by $\rho_{tt}$ 
and $\rho_{tc}$~\cite{Fuyuto:2017ewj}, 
but it is unclear which way to go. 
%{\color{red}So we should be open to the possibility that both extra $\tau$ ($\rho_{\tau\mu}$ and $\rho_{\tau\tau})$ and extra top ($\rho_{tc} \ \rm and \ \rho_{tt} $) Yukawa coupling can drive EWBG}. 
%
Nonzero $\rho_{tc}$ and $\rho_{\tau\mu}$ opens up 
some exciting new channels, such as 
$cg\to tH, tA \to t\tau\mu$, $cg\to bH^+ \to bHW^+
 \to bW^+\tau\mu$~\cite{Hou:2019grj}, 
hence providing a rich phenomenology for LHC.

In this article, we find that beyond $2m_t$ 
it is quite challenging to probe either 
$H,A \to \tau\mu$ or $H,A \to \tau \tau$. 
However, we have to keep in mind that we have 
not considered all $\tau$-lepton decay modes, 
so by combining all channels of $\tau$ decay and 
performing more sophisticated machine learning 
classifications, the combined channels may hold 
promising future for discovering $H$ and $A$ 
at the HL-LHC, or even future FCC-hh or SppC colliders. 

Although we have not focused on the case in this article,
let us end with a positive note by connecting 
$pp \to H,\,A \to \tau\mu$ search with the 
recent confirmation of the muon $g-2$ anomaly. 
In g2HDM, the muon $g-2$ anomaly can be accounted for by 
sizable $\rho_{\tau\mu}\rho_{\mu\tau}$ 
and relatively low mass $H$ or $A$.
If one takes the muon $g-2$ anomaly seriously,
it may well mean that CMS $pp \to H,\,A \to \tau\mu$ 
search might draw a hint of a signal 
below the $2m_t$ threshold with full Run~2 data.

\vskip0.2cm
{\noindent \bf Acknowledgements:} WSH and RJ are 
supported by MOST 110-2639-M-002-002-ASP of Taiwan, 
with WSH in addition by NTU 111L104019 and 111L894801.
CK is supported by the University of Oklahoma.

%relevant papers (contraint section) : Belle, CMS, ATLAS results, Alignment papers, 2HDMC manual, landau ghost issue,  B-B mixing , B-Physics, h->gammagamma issue, charged Higgs studies arguments

%\clearpage

%============================================================
% END DOCUMENT
%============================================================                  
\end{document}